\documentclass[12pt]{article}

\usepackage{preprintstyle}
\usepackage[utf8]{inputenc}

\newcommand{\beq}{\begin{equation}}
\newcommand{\eeq}{\end{equation}}
\newcommand{\bal}{\begin{aligned}}
\newcommand{\eal}{\end{aligned}}
\newcommand{\rmd}{\mathrm d}

\usepackage{physics, xparse}
\usepackage{dsfont}
\usepackage[normalem]{ulem}
\usepackage{cancel,xcolor}

\title{Quantum Extremal Modular Curvature: Modular Transport with Islands}

\author{Lars Aalsma$^{a,b}$,}
\author{Cynthia Keeler$^{b}$,}
\author{Claire Zukowski$^{c}$,}
\emailAdd{laalsma@asu.edu}
\emailAdd{keelerc@asu.edu}
\emailAdd{czukowsk@d.umn.edu}
\affiliation{$^a$Beyond: Center for Fundamental Concepts in Science, Arizona State University, Tempe, Arizona 85287, USA}
\affiliation{$^b$Department of Physics, Arizona State University, Tempe, Arizona 85287, USA}
\affiliation{$^c$Department of Physics and Astronomy, University of Minnesota Duluth, Duluth, MN 55812, USA}

\abstract{Modular Berry transport is a useful way to understand how geometric bulk information is encoded in the boundary CFT: The modular curvature is directly related to the bulk Riemann curvature. We extend this approach by studying modular transport in the presence of a non-trivial quantum extremal surface. Focusing on JT gravity on an AdS background coupled to a non-gravitating bath, we compute the modular curvature of an interval in the bath in the presence of an island: the Quantum Extremal Modular Curvature (QEMC). We highlight some important properties of the QEMC, most importantly that it is non-local in general. In an OPE limit, the QEMC becomes local and probes the bulk Riemann curvature in regions with an island. Our work gives a new approach to probe physics behind horizons.}

\begin{document}

\maketitle

\section{Introduction}
In the context of the AdS/CFT correspondence, entanglement entropy has provided key insights as to how geometry emerges from conformal field theories. The Ryu-Takayanagi (RT) formula \cite{Ryu:2006bv} (and its covariant generalization \cite{Hubeny:2007xt}) has played an essential role in these developments as it relates CFT entanglement entropy to the area of bulk minimal surfaces. Of course, the RT formula is just one example that demonstrates how some piece of geometric information is encoded in the CFT. A promising approach towards understanding how more general geometric properties emerge from holographic CFTs is modular transport.

This method, pioneered in \cite{Czech:2017zfq, Czech:2019vih} and inspired by related work on kinematic space \cite{Czech:2015qta, deBoer:2015kda, Czech:2016xec, deBoer:2016pqk}, defines a parallel transport problem for modular Hamiltonians that are deformed by changing CFT subregions (see also \cite{Nogueira:2021ngh, Banerjee:2022jnv} for other approaches to Berry transport in holography, and \cite{Bao:2019bib,Engelhardt:2016wgb,Engelhardt:2016crc} for other methods for reconstructing the bulk geometry). The transport process admits a redundancy in the form of zero modes that commute with the modular Hamiltonian. This redundancy is analogous to the Berry phase in quantum mechanics and was therefore referred to as a modular Berry phase. By gauging this redundancy one finds a modular curvature, again analogous to the well-known Berry curvature. One key result of \cite{Czech:2019vih} was to show how---in holography---geometry is encoded in the modular curvature. In many cases of interest, the action of the modular Hamiltonian reduces to a boost isometry near the edge of the entanglement wedge. Exploiting this relation, it was demonstrated that the modular curvature is related to the Riemann tensor \cite{Czech:2019vih,DeBoer:2019kdj}.

Subsequent work not only considered changes in the modular Hamiltonian due to changing CFT subregions (known as shape deformations), but also changes due to modifying the state (referred to as state deformations) \cite{deBoer:2021zlm,Czech:2023zmq}. The references \cite{deBoer:2021zlm,Czech:2023zmq} showed that the modular curvature associated to state deformations is related to a bulk symplectic form. What these different examples have in common is that the modular curvature probes some piece of interesting bulk geometric information.

When considering a subregion in the CFT, we only have access to the portion of the bulk covered by the entanglement wedge \cite{Dong:2016eik}. So-called ``entanglement shadow regions'' were shown to be impenetrable by boundary-anchored minimal surfaces \cite{Balasubramanian:2014sra, Freivogel:2014lja}, while surfaces of nonpositive extrinsic curvature dubbed ``barriers'' were shown to obstruct boundary-anchored extremal surfaces in the covariant description \cite{Engelhardt:2015dta}. Since the modular transport approach to bulk reconstruction uses bulk extremal surfaces to probe the Riemann curvature, it has the seeming drawback that barriers may prevent reconstruction in arbitrary bulk regions, for instance deep inside a black hole.

One motivation of this work is to see how this obstruction can be partly overcome by making use of quantum extremal surfaces (QESs) \cite{Engelhardt:2014gca}. Studies on the black hole information paradox revealed that the entanglement wedge of black hole radiation can include an `island' QES region that is disconnected from the boundary and can lie behind the horizon \cite{Penington:2019npb,Almheiri:2019psf}. In this paper, we therefore focus on modular transport in the presence of an island which involves performing modular flow on two disjoint subregions. We are interested in understanding if and what geometrical information about the bulk is encoded in the modular curvature in the presence of a non-trivial QES: a quantity which we refer to as the `Quantum Extremal Modular Curvature' (QEMC). 

We focus on a $1+1$-dimensional free fermion theory coupled to JT gravity, since there is a closed-form expression for its modular Hamiltonian for $n$ disjoint subregions \cite{Casini:2009vk}. The modular Hamiltonian of the free fermion was used in the context of islands in \cite{Chen:2019iro}, and its modular curvature on flat space backgrounds has been studied in \cite{Chen:2022nwf}. Our contribution is to put these results to work in our setup of interest, which is JT gravity on an AdS$_2$ background coupled to a non-gravitating bath region. The free fermion theory lives on this background and allows us to analytically study modular transport in the presence of islands.

Our setup is somewhat different than the one considered in previous studies of modular transport in holography. Typically, a subregion in the CFT$_d$ is used to define bulk entanglement wedges in AdS$_{d+1}$: modular flow in the CFT$_d$ has a direct analog in AdS$_{d+1}$. However, the dual of JT gravity coupled to a non-gravitational bath is a quantum mechanical (QM) theory. Instead of defining subregions in this QM theory we work directly in the semi-classical `gravitational picture' and define a subregion in the two-dimensional non-gravitational bath which, via the QES condition, also identifies a subregion in the bulk. This is the same as the setup for computing the entanglement entropy of Hawking radiation in the presence of islands.

A main result of our paper is that---in an OPE limit---the QEMC still probes bulk Riemann curvature including near the island region. This yields a method to access geometrical information, possibly behind a horizon, from the non-gravitational bath region. Away from this simplifying limit, the curvature contains non-local terms that have their origin in the non-local nature of the modular Hamiltonian for disjoint intervals. It is tempting to speculate that these non-local corrections signal some sort of breakdown of the semi-classical EFT along the lines of \cite{Bousso:2023kdj,Banks:2024imv}, but we leave further interpretation to future work.

The rest of this paper is organized is follows. In Sec. \ref{sec:ModTranRev}, we review the process of modular transport and explain how it probes bulk Riemann curvature. As a concrete and illustrative example we focus on AdS$_3$/CFT$_2$, generalizing some of the results in \cite{deBoer:2021zlm,Czech:2023zmq}. In Sec. \ref{sec:ModTranJT}, we focus on JT gravity and consider a solution that contains a non-trivial quantum extremal surface. We then introduce modular flow in this background and compute various components of the QEMC. We end with a discussion of our results in Sec. \ref{sec:Discussion}.

\section{Modular Transport and Berry Curvature} \label{sec:ModTranRev}
We start with a short review of modular Hamiltonians and the transport process that probes the bulk geometry.

\subsection{Introduction to Modular Transport}
Given a quantum state $\ket{\psi}$ and a spatial subregion $A$ the reduced density matrix is defined as
\beq
\rho_A = \text{tr}_{\bar A}\ket{\psi}\bra{\psi} ~,
\eeq
where we perform a trace over $\bar A$, the complement of $A$. The formal definition of the modular Hamiltonian $H$ of this subregion is
\beq
H = -\ln\rho_A ~.
\eeq
Using $s$ to denote the modular `time' parameter, the evolution of an operator under modular flow is given by
\beq
{\cal O}(s) \to {\cal O}(s') = e^{\frac {i(s'-s)}{2\pi}H}{\cal O}(s)e^{-\frac {i(s'-s)}{2\pi}H} ~.
\eeq

\begin{figure}[t]
\centering
\includegraphics[scale=.9]{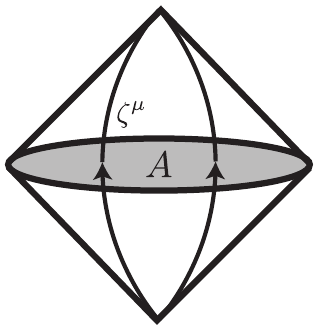}
\caption{Causal diamond associated to the subregion $A$. The vector field $\zeta^\mu$ is a boost vector that preserves the diamond.}
\label{fig:diamond}
\end{figure}

For general states and subregions, the modular Hamiltonian is a complicated and possibly non-local operator. However, in certain special cases such as a single interval in the Minkowski vacuum \cite{Bisognano:1976za}, it obeys the local expression 
\beq \label{eq:localModHam}
H = \int_A \rmd \Sigma \,\zeta^\mu n^\nu T_{\mu\nu} ~.
\eeq
Here $\zeta^\mu$ is the boost vector that preserves the causal diamond defined by the subregion $A$ (see Figure \ref{fig:diamond}),  $n^\mu$ is the unit normal to $A$, and $T_{\mu\nu}$ is the stress tensor.

As explained in \cite{Faulkner:2017vdd}, locality of the modular Hamiltonian is directly related to ultralocality of correlation functions, in the vacuum. More generally, for states that are perturbations away from the vacuum, the modular Hamiltonian acts as a local boost close to the horizons of the diamond \cite{Balakrishnan:2020lbp}. Intuitively, excited states like those considered in \cite{Balakrishnan:2020lbp} look like the vacuum at short distances. For multiple subregions in the vacuum, the modular Hamiltonian again is non-local.

When changing the subregion or the state, the modular Hamiltonian also changes. As studied in depth in \cite{Czech:2019vih,DeBoer:2019kdj,deBoer:2021zlm,Czech:2023zmq}, we can define a modular transport problem by parallel transporting the modular Hamiltonian. In holographic setups, modular transport is a useful probe of local bulk physics. To define this transport problem, we consider a family of nearby modular Hamiltonians $H(\lambda)$ labeled by $\lambda$. Assuming we can diagonalize the Hamiltonian using a unitary vector $U(\lambda)$ as
\beq
D(\lambda) = U(\lambda)H(\lambda)U^\dagger(\lambda) ~,
\eeq
then a change in the modular Hamiltonian can be expressed as:
\beq \label{eq:ModTransEq}
\partial_\lambda H = [\partial_\lambda U^\dagger U,H] + U^\dagger\partial_\lambda D U ~.
\eeq
The second term on the right-hand side commutes with the modular Hamiltonian and is known as a modular zero mode. We can explicitly represent the zero modes as $O=e^{i\sum_a c_aQ_a}$, where $c_a$ are constant coefficients and $[Q_a,H]=0$. 

Transforming the modular Hamiltonian by a zero mode as
\beq
\tilde H = \tilde U^\dagger D \tilde U ~,
\eeq
with $\tilde U = UO$, leaves \eqref{eq:ModTransEq} invariant. Thus the choice of zero mode frame is a gauge redundancy with an associated connection $\Gamma:=P_0[\partial_\lambda U^\dagger U]$, where $P_0$ is a projector onto the zero mode. If we transform $U$ by a zero mode $U\to\tilde U = UO$, the connection changes as
\beq
\Gamma\to\tilde \Gamma = P_0[O^\dagger \Gamma O - O^\dagger\partial_\lambda O] =P_0[ -O^\dagger {\cal D}_\lambda O] ~,
\eeq
where ${\cal D}_\lambda$ is a covariant derivative given by
\beq
 {\cal D}_\lambda O = (\partial_\lambda - \Gamma)O ~.
\eeq
If we fix a gauge by setting $\tilde \Gamma=0$ we obtain
\beq
{\cal D}_\lambda O = (\partial_\lambda - \Gamma)O = 0 ~.
\eeq
Dropping the tildes, this leads to the following modular transport equations:
\beq
\bal
\partial_\lambda H - P_0[\partial_\lambda H] &=   [\partial_\lambda U^\dagger U,H] \\
P_0[\partial_\lambda U^\dagger U] &=0 ~.
\eal
\eeq
Considering different connections obtained by different variations and projecting out the zero mode, the modular transport equations are solved by 
\beq
V_i = (1-P_0)\partial_{\lambda_i}U^\dagger U ~.
\eeq
We refer to $V_i$ as the generator of modular transport. The modular curvature is then given by~\cite{deBoer:2021zlm}\footnote{In~\cite{deBoer:2021zlm}, the general expression for the curvature also has an overall zero mode projector. We choose to omit this, since there are no non-zero mode contributions in the examples we consider in this paper.}
\beq
{\cal R}_{ij} := [V_i,V_j] ~.
\eeq

\subsection{Modular Transport in Holography}
\begin{figure}[t]
\centering
\includegraphics[scale=1.3]{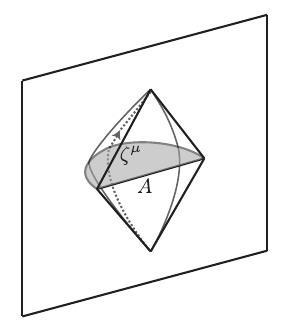}
\caption{An interval $A$ in a two dimensional CFT defines a causal diamond on the AdS boundary. The entanglement wedge of $A$ defines the bulk subregion where the bulk modular Hamiltonian acts. Near the bulk entangling surface the modular Hamiltonian reduces to the (bulk) boost Killing vector $\zeta^\mu$.}
\label{fig:EntWedge}
\end{figure}

In AdS$_{d+1}$/CFT$_d$, modular transport is a probe of the bulk geometry from a boundary perspective. The relation between the bulk and boundary modular Hamiltonian is given by the JLMS formula \cite{Jafferis:2015del},
\beq \label{eq:JLMS}
H_{\rm bdy} = \frac{\hat A}{4G_{d+1}} + H_{\rm bulk} + {\cal O}(G_{d+1})~.
\eeq
$G_{d+1}$ is the $(d+1)$-dimensional Newton constant and $\hat A$ is the area operator of the bulk entanglement wedge associated to the subregion in the CFT. As mentioned before, when we consider vacuum states or take the limit of being close to the entangling surface, modular flow reduces to the boost Killing vector associated to the causal diamond of the subregion. We've drawn the example of AdS$_3$/CFT$_2$ in Figure \ref{fig:EntWedge}.

As explained in detail in \cite{Czech:2019vih} (see also \cite{DeBoer:2019kdj,deBoer:2021zlm,Czech:2023zmq}), we can interpret the zero mode redundancy of the modular Hamiltonian in the bulk as a choice of a particular zero mode frame attached to an entanglement wedge. Comparing relative zero mode frames for different entanglement wedges then gives us information about the bulk curvature. Accordingly, the modular curvature can be used to directly obtain information about the bulk Riemann curvature. \cite{Czech:2019vih,DeBoer:2019kdj} provides a formal derivation of this relationship. Our example in the context of AdS$_3$/CFT$_2$ is (hopefully) instructive to see explicitly how the Riemann curvature appears in the modular curvature.

\subsubsection*{Modular Transport in CFT$_2$}
We now show how to obtain the bulk Riemann curvature using modular transport in AdS$_3$/CFT$_2$. We first consider the modular Hamiltonian and transport in a two-dimensional CFT, and then construct the dual quantities in pure AdS$_3$, following \cite{Czech:2019vih}.

For a two-dimensional CFT in the vacuum, the modular Hamiltonian can be constructed from the generators of the conformal algebra. We put the CFT on a cylinder and work with lightcone coordinates $x^\pm = \tau \pm \theta$, where $\tau$ is the time parameter and $\theta$ is the angle on the cylinder. The modular Hamiltonian decomposes into a left-moving and right-moving piece:
\beq
H = H_{(+)} + H_{(-)} ~.
\eeq
Defining a conformal Killing vector (CKV) as $\zeta  = \zeta^+\partial_+ + \zeta^-\partial_-$ the modular Hamiltonian is given by
\beq \label{eq:IntRepModHam}
H_{(\pm)} = 2\pi\int \rmd x^\pm\, \zeta^\pm T_{\pm\pm} ~,
\eeq
where we choose a normalization of $2\pi$ in front. We use brackets in the subscript to avoid confusion with the vector components. Explicitly, we have
\beq \label{eq:CKV-CFT}
\zeta^\pm = s_1^\pm L^\pm_1 + s^\pm_0L_0^\pm + s_{-1}^\pm L_{-1}^\pm ~,
\eeq
where the components of the generators $L_N = L_N^+\partial_+ + L_N^-\partial_-$ can be compactly written as
\beq \label{eq:ConfGen}
L_N^\pm = e^{iN x^\pm} ~.
\eeq 
Here, $N$ runs over $0, \pm 1$. The generators are CKVs and therefore obey
\beq
{\cal L}_{L_N} g_{\mu\nu} - (\partial_\rho L^\rho_N) g_{\mu\nu} = 0 ~.
\eeq

We now specify an interval in the CFT. We label the left endpoint by $x^\pm_i = a^\pm$ and the right endpoint by $x_f^\pm = b^\pm$. Next, we construct the modular Hamiltonian of this interval by finding the CKV that preserves the causal diamond, i.e. we want the norm of the CKV to vanish at the horizons. This vector is given by \eqref{eq:CKV-CFT} with the coefficients
\beq
\bal \label{eq:coefficients}
s_1^\pm &= \pm \frac{\cot\left[\frac{b^\pm-a^\pm}{2}\right]}{e^{ia^\pm}+e^{ib^\pm}} ~, \\
s_0^\pm &= \mp\cot\left[\frac{b^\pm-a^\pm}{2}\right] ~, \\
s_{-1}^\pm &= \pm\frac{\cot\left[\frac{b^\pm-a^\pm}{2}\right]}{e^{-ia^\pm}+e^{-ib^\pm}} ~.
\eal
\eeq

By taking derivatives of the modular Hamiltonian, we can construct the generators of modular transport. Instead of working with the integral representation \eqref{eq:IntRepModHam}, we opt to directly use the CKV
\beq
K:=2\pi\zeta ~.
\eeq
The generators of modular transport are then given by
\beq \label{eq:ModGen1}
V_{a^\pm} = \pm\frac1{2\pi}\partial_{a^\pm}K ~, \quad V_{b^\pm} = \mp\frac1{2\pi}\partial_{b^\pm}K ~,
\eeq
and it is straightforward to check that they solve the modular transport equations:
\beq \label{eq:ModGen2}
[V_{a^\pm},K] = \partial_{a^\pm}K ~, \quad [V_{b^\pm},K] = \partial_{b^\pm}K ~.
\eeq
There is no zero mode that needs to be projected out, as there is no part of $\partial_\lambda K$ that commutes with $K$.

Just as in the integral representation of the modular Hamiltonian, we find it convenient to split $K = K_{(+)}+K_{(-)}$ where
\beq
K_{(+)} := 2\pi \xi^+\partial_+ ~, \quad K_{(-)} := 2\pi \xi^-\partial_- ~.
\eeq
In two dimensions, this notation is rather redundant since it just labels the $\pm$ components of $K$. However, next when we consider AdS$_3$ $K_{(\pm)}$ will also involve the radial direction. The non-zero components of the modular curvature are now given by
\beq
\bal \label{eq:ModCurvCFT}
{\cal R}_{a^+b^+} = [V_{a^+},V_{b^+}] = +\frac1{2\sin^2\left[\frac{a^+-b^+}{2}\right]}\frac{K_{(+)}}{2\pi} ~,\\
{\cal R}_{a^-b^-} = [V_{a^-},V_{b^-}] = -\frac1{2\sin^2\left[\frac{a^--b^-}{2}\right]}\frac{K_{(-)}}{2\pi} ~.
\eal
\eeq

We use $\lambda= (a^+,a^-,b^+,b^-)$ to refer to the four parameters that define our interval.  We then consider an infinitesimal change $\lambda' = \lambda+\rmd\lambda$.  We can parameterize all possible changes via the following basis:
\beq
\bal \label{eq:BasisShift}
\lambda_1 &= (a^+-\rmd\lambda,a^--\rmd\lambda,b^+-\rmd\lambda,b^--\rmd\lambda) ~,\\
\lambda_2 &= (a^+-\rmd\lambda,a^-+\rmd\lambda,b^+-\rmd\lambda,b^-+\rmd\lambda) ~,\\
\lambda_3 &= (a^+-\rmd\lambda,a^--\rmd\lambda,b^+-\rmd\lambda,b^-+\rmd\lambda) ~,\\
\lambda_4 &= (a^+-\rmd\lambda,a^--\rmd\lambda,b^++\rmd\lambda,b^--\rmd\lambda) ~.
\eal
\eeq
The generator of modular transport for any infinitesimal deformation is a linear combination of the four generators
\beq \label{eq:ModGen3}
V_{\lambda_i} = \left(\frac{\partial a^+}{\partial \lambda_i}\right)V_{a^+} + \left(\frac{\partial a^-}{\partial \lambda_i}\right)V_{a^-} + \left(\frac{\partial b^+}{\partial \lambda_i}\right)V_{b^+} + \left(\frac{\partial b^-}{\partial \lambda_i}\right)V_{b^-}~.
\eeq

\begin{figure}[t]
\centering
\includegraphics[scale=.6]{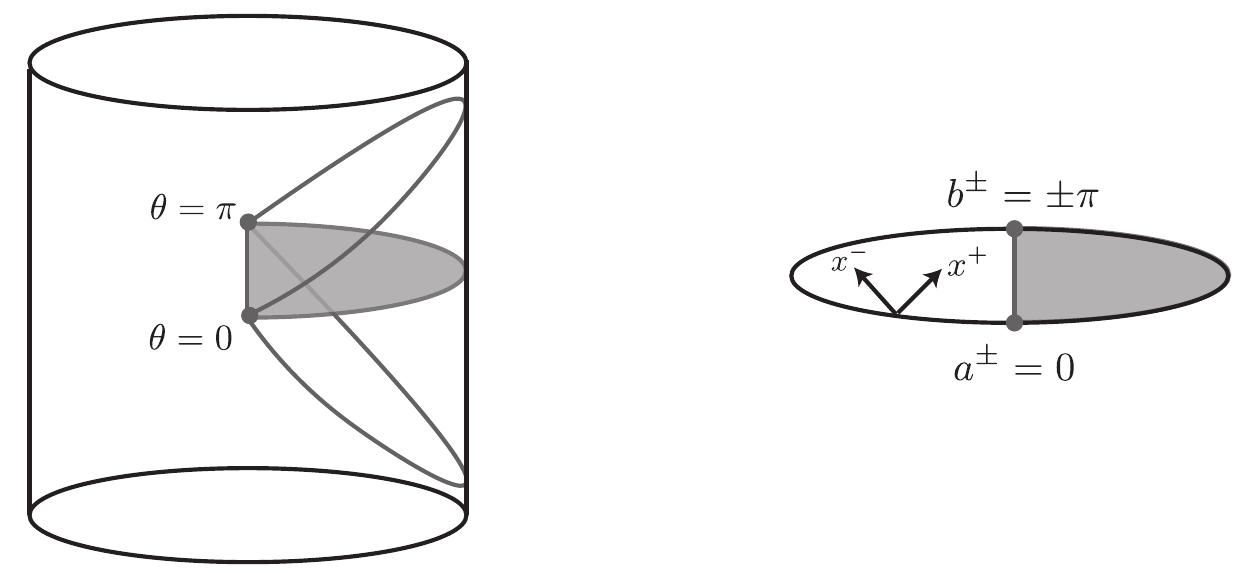}
\caption{Global AdS$_3$ corresponds to a cylinder. We choose an interval in the CFT$_2$ that covers half of the cylinder. The Ryu-Takayanagi surface is then given by a simple diagonal geodesic.}
\label{fig:Cylinder}
\end{figure}

We now specify to an interval given by $\tau=0$ and $\theta=[0,\pi]$ which covers half of the cylinder. This choice has the advantage that the minimal surface in the bulk corresponds to a simple diagonal geodesic, see Figure \ref{fig:Cylinder}.
In lightcone coordinates the interval is given by $[a^\pm,b^\pm] = [0,\pm \pi]$. For this choice of interval the generators for the basis \eqref{eq:BasisShift} become
\beq
\bal
V_{\lambda_1} &= \partial_+ + \partial_- ~, \\
V_{\lambda_2} &= \partial_+ - \partial_- ~, \\
V_{\lambda_3} &= \partial_+ + \cos(x^-)\partial_- ~, \\
V_{\lambda_4} &= \cos(x^+)\partial_+ - \partial_- ~.
\eal
\eeq
The first two generators have an especially simple interpretation as they correspond to a translation in the $\tau$ and $\theta$ direction respectively.

\subsubsection*{Modular Transport in AdS$_3$}
We now turn to the bulk dual of the modular transport problem described in the two-dimensional CFT. We consider the AdS$_3$ metric in global coordinates,
\beq
\rmd s^2 = \ell^2(-\cosh^2\rho \,\rmd \tau^2 + \rmd\rho^2 + \sinh^2\rho\,\rmd\theta^2) ~.
\eeq
Lightcone coordinates in the bulk are similarly defined as $x^\pm = \tau \pm \theta$. The AdS boundary is located at $\rho\to\infty$.

The bulk modular Hamiltonian can be found by extending the CKV's of the causal diamond in the boundary to the bulk. In fact, the conformal algebra satisfied by \eqref{eq:ConfGen} corresponds to the algebra satisfied by the Killing vectors (KVs) of AdS$_3$. The CKV in the CFT$_2$ is therefore the boundary value of the full bulk KV in AdS$_3$ that preserves the entanglement wedge.

With a slight abuse of notation, we again write $K=2\pi\zeta$. $K$ splits into $K_{(+)}+K_{(-)}$ which are given by
\beq \label{eq:K-AdS}
K_{(\pm)} = 2\pi\left(s^\pm_{1}J_1^\pm + s^\pm_{0}J_0^\pm + s^\pm_{-1}J_{-1}^\pm\right) ~,
\eeq
with
\beq
\bal
J^\pm_1 &= e^{ix^\pm}\coth(2\rho)\partial_\pm - e^{ix^\pm}\csch(2\rho)\partial_\mp - \frac i2 e^{ix^\pm}\partial_\rho  ~, \\
J^\pm_0 &= \partial_\pm ~, \\
J^\pm_{-1} &= \frac12e^{-ix^\pm}(\coth\rho+\tanh\rho)\partial_\pm - e^{-ix^\pm}\csch(2\rho)\partial_\mp + \frac i2e^{-ix^\pm}\partial_\rho ~.
\eal
\eeq
It is straightforward to check that $J^\pm_N$ are KVs of AdS$_3$ and satisfy
\beq
\left.\lim_{\rho\to\infty}J^\pm_N \right|_{\partial_\rho = 0} = L^\pm_{N} ~.
\eeq

We want the KV to preserve the entanglement wedge, thus its norm needs to vanish at the horizons. This requirement is equivalent to the requirement that the CKV in the CFT$_2$ preserves the causal diamond, so the coefficients are again given by \eqref{eq:coefficients}.

Again defining the generators of modular transport $V_{a^\pm},\, V_{b^\pm}$ as in \eqref{eq:ModGen1}, but with $K$ as in \eqref{eq:K-AdS}, as before we find the modular transport equations \eqref{eq:ModGen2}. Additionally we can use the same basis \eqref{eq:ModGen3} to express a general infinitesimal deformation. Similarly, the modular curvature components are given by \eqref{eq:ModCurvCFT}. Explicitly:
\beq
\bal
{\cal R}_{a^+b^+} = [V_{a_+},V_{b^+}] = +\frac1{2\sin^2\left[\frac{a^+-b^+}{2}\right]}\frac{K_{(+)}}{2\pi} ~,\\
{\cal R}_{a^-b^-} = [V_{a_-},V_{b^-}] = -\frac1{2\sin^2\left[\frac{a^--b^-}{2}\right]}\frac{K_{(-)}}{2\pi} ~.
\eal
\eeq
Now, $K_{(\pm)}$ refers to \eqref{eq:K-AdS}.

As before, choosing the diagonal geodesic $(a^\pm,b^\pm) = (0,\pm\pi)$, the entangling surface in the bulk is given by the diagonal geodesic that connects these two endpoints (as we displayed in Figure \ref{fig:Cylinder}). For this geodesic, the generators in our basis take the form
\beq
\bal
V_{\lambda_1} &= \partial_+ + \partial_- ~, \\
V_{\lambda_2} &= \partial_+ - \partial_- ~, \\
V_{\lambda_3} &= (1-\cos x^-\csch(2\rho))\partial_+ + \cos x^-\coth(2\rho)\partial_- +\frac12\sin x^-\partial_\rho ~, \\
V_{\lambda_4} &= \cos x^+\coth(2\rho)\partial_+ + (1-\cos x^+\csch(2\rho))\partial_- +\frac12\sin x^+\partial_\rho ~.
\eal
\eeq

\subsubsection*{Relation between Modular and Riemann Curvature}
As explained in \cite{Czech:2019vih}, the modular curvature is a probe of the Riemann curvature in the bulk. Since the modular Hamiltonian implements a boost isometry that preserves the entanglement wedge (at least near its edge), it is perhaps unsurprising that we can probe the bulk geometry by studying how this Hamiltonian changes under modular transport.

For modular Hamiltonians that reduce to a boost only near the edge of the entanglement wedge, it is convenient to set up a Riemann normal coordinate system in which the metric is flat at the boundary of the entanglement wedge. In this coordinate system, \cite{DeBoer:2019kdj} derived the relation between the modular curvature and the Riemann tensor.\footnote{See (3.20) of \cite{DeBoer:2019kdj}. It should be noted however that this expression contains a typo that we correct in Appendix \ref{sec:FNC}.} We demonstrate this method to obtain the Riemann curvature for three-dimensional theories using Fermi normal coordinates in Appendix \ref{sec:FNC}.

In this paper we will also consider modular flow for subregions that are not attached to the boundary. For a single subregion that is not attached to the boundary the modular Hamiltonian is not given by a Killing vector, but a conformal Killing vector.

We will only study situations where the modular Hamiltonian is an exact (conformal) Killing vector. In this case, the relation between the modular curvature and Riemann curvature follows more directly. It is no longer necessary to make use of a Riemann normal coordinate system. Accordingly, we define the modular Hamiltonian $K$ of the entanglement wedge associated to a CFT$_2$ interval with endpoints $(a^\pm,b^\pm)$. We then consider an infinitesimal displacement of the entanglement wedge in the $+$ direction described by the generators $V_{a^+}$ and $V_{b^+}$. For convenience, we define 
\beq
\bal
K'_{a^+} \equiv \frac{1}{(2\pi)^2} \partial_{a^+} K &= \frac{1}{(2\pi)^2}[V_{a^+},K] ~, \\
K'_{b^+} \equiv \frac{1}{(2\pi)^2} \partial_{b^+} K&= \frac{1}{(2\pi)^2}[V_{b^+},K] ~.
\eal
\eeq
Absorbing the factors of $2\pi$ here results in a nice looking expression for the modular curvature below.

Using the definition of the generators \eqref{eq:ModGen2}, we can express
\beq
V_{a^+} = [K'_{a^+},K] ~, \quad V_{b^+} = -[K'_{b^+},K] ~.
\eeq
By working out the commutators, the modular curvature can now be expressed as
\beq
\bal
({\cal R}_{a^+b^+})^\mu = [V_{a^+},V_{b^+}]^\mu &= V^\nu_{a^+}\left(2R_{\sigma\nu\rho}^{~~~~\mu}K'{}_{b^+}^{[\rho}K^{\sigma]}+\nabla_\nu K'{}_{b^+}^\rho\nabla_\rho K^\mu-\nabla_\nu K^\rho\nabla_\rho K'{}_{b^+}^\mu\right) \\
&-\nabla_\nu V_{a^+}^\mu V_{b^+}^\nu ~.
\eal
\eeq
The appearance of the Riemann tensor is a direct consequence of the fact that $K$ and $K'_{b^+}$ are Killing vectors. This follows from the fact that for a general Killing vector $Y$ we can use the identity $\nabla_\mu\nabla_\nu Y^\rho = R_{\sigma\mu\nu}^{~~~~\rho}Y^\sigma$, assuming the metric is torsion-free. This expression therefore gives a relation between the modular curvature and bulk Riemann curvature in addition to terms involving the (change of the) modular Hamiltonian.

Later on, we will study modular transport for an interval in a $1+1$-dimensional gravitational theory. In that case, the modular Hamiltonian is given by a CKV. For that purpose, we also denote the relation when instead of Killing vectors, we only have CKVs. In that case we obtain an additional term $\Omega^\mu$ (in $d+1$-spacetime dimensions):
\beq \label{eq:ConformalCurvature}
\bal
[V_{a^+},V_{b^+}]^\mu &= V^\nu_{a^+}\left(2R_{\sigma\nu\rho}^{~~~~\mu}K'{}_{b^+}^{[\rho}K^{\sigma]}+\nabla_\nu K'{}_{b^+}^\rho\nabla_\rho K^\mu-\nabla_\nu K^\rho\nabla_\rho K'{}_{b^+}^\mu\right) \\
&-\nabla_\nu V_{a^+}^\mu V_{b^+}^\nu +\frac1{d+1} \Omega^\mu ~.
\eal
\eeq
$\Omega^\mu$ is defined as
\beq
\Omega^\mu = V_{a^+}^\nu\big[K^\rho\nabla^\mu(\nabla\cdot K'{}_{b^+})g_{\nu\rho} - K^\rho\nabla_\rho(\nabla\cdot K'{}_{b^+})g_\nu^{\,\,\,\mu}-K^\rho\nabla_\nu(\nabla\cdot K'{}_{b^+})g_\rho^{\,\,\,\mu} - \left(K \leftrightarrow K'{}_{b^+}\right)\big] ~,
\eeq
and vanishes when $K$ and $K'{}_{b^+}$ are Killing vectors.\footnote{To derive this relation, we used the following identity for a CKV $\zeta$, which can be derived from the definition of curvature from parallel transport of a vector field combined with the algebraic Bianchi identity:
\beq
\nabla_\mu\nabla_\nu\zeta^\rho = R_{\sigma \mu \nu}^{\ \ \ \ \rho}\zeta^\sigma - \frac1d\left(\nabla^\rho(\nabla\cdot\zeta)g_{\mu\nu}-\nabla_\nu(\nabla\cdot\zeta)g_\mu^{\ \rho}-\nabla_\mu(\nabla\cdot\zeta)g_{\nu}^{\ \rho}\right) ~,
\eeq
again assuming the metric is torsion-free.}

This finishes our discussion of modular transport in the context of AdS$_3$/CFT$_2$ and the relation between the modular curvature and the Riemann curvature. Next, we specialize to a (1+1)-dimensional bulk and consider modular transport in JT gravity.

\section{Modular Transport in JT Gravity} \label{sec:ModTranJT}
Now that we have seen how modular transport probes bulk curvature, we shift our focus to two-dimensional JT gravity. This choice has the advantage that we can incorporate the effect of quantum extremal surfaces and see---in the presence of an island---how modular transport probes geometry and entanglement of disconnected entanglement wedges. In certain simplifying limits, modular flow becomes local and the modular curvature is directly related to the Riemann curvature of the quantum extremal surface. However, in general the `Quantum Extremal Modular Curvature' is a non-geometric object that captures non-local effects in the entanglement structure.

First, we start with a review of JT gravity.

\subsection{JT Gravity on AdS$_2$}
We now review a two-dimensional black hole solution in JT gravity. We start with the action of JT gravity on an AdS$_2$ background:
\beq
I = \frac1{16\pi G_2}\int \rmd ^2x\sqrt{-g}\left(\phi R + \frac2{\ell^2}(\phi-\phi_0)\right) + I_{\rm CFT} + I_{\rm bdy} ~.
\eeq
Here $I_{\rm CFT}$ corresponds to a coupling to a CFT and $I_{\rm bdy}$ contains a boundary term. The equations of motion are
\beq
\bal
 - \nabla_\mu\nabla_\nu\phi + g_{\mu\nu}\square\phi - \frac{(\phi-\phi_0)}{\ell^2}g_{\mu\nu} - 8\pi G_2 T_{\mu\nu} &= 0 ~,\\
R+2/\ell^2 &= 0 ~.
\eal
\eeq
If we are interested in solutions where the dilaton exhibits a symmetry, then it is easiest to work in coordinates that preserve the same symmetry. In our case, we focus on time-independent solutions so it is appropriate to consider the AdS$_2$ metric in Poincaré coordinates:
\beq
\rmd s^2 = \frac{\ell^2}{x^2}(-\rmd t^2 + \rmd x^2) ~.
\eeq
We take $x\leq 0$ with $x=0$ the AdS boundary. With this choice the non-gravitating bath region that we couple to is located at $x>0$. This metric has a timelike Killing vector and imposing the dilaton to be invariant under this symmetry, the vacuum solution (i.e. $T_{\mu\nu}=0$) for the dilaton is given by
\beq
\phi = \phi_0 -\frac{\phi_r}{x} ~,
\eeq
where $\phi_0$ and $\phi_r$ are constants.

Because left-movers and right-movers decouple in two dimensions it is convenient to introduce Kruskal coordinates 
\beq
x^\pm = t \pm x ~,
\eeq
in which the metric and dilaton solution become
\beq
\bal \label{eq:AdSmetric}
\rmd s^2 &= - \frac{4\ell^2}{(x^+-x^-)^2}\rmd x^+\rmd x^- ~, \\
\phi &= \phi_0 - \frac{2\phi_r}{x^+-x^-} ~.
\eal
\eeq

Now we allow the stress tensor to be nonzero. First, we treat this matter coupling classically. Conformal invariance then implies that the trace of the stress tensor vanishes. In this case, the equations of motion can be integrated to obtain the general solution \cite{Almheiri:2014cka,Engelsoy:2016xyb}
\beq \label{eq:DilSol}
\phi = \phi_0 -\frac{2\phi_r}{x^+-x^-} + 8\pi G_2\left(I_+ - I_- \right) ~.
\eeq
Here we defined the two functions
\beq
\bal
I_+ &= \int^{x^+}_{x^+_0} \rmd s \frac{(x^+-s)(x^--s)}{(x^+-x^-)}T_{++}(s) ~,\\
I_- &= \int_{x^-_0}^{x^-} \rmd s \frac{(x^+-s)(x^--s)}{(x^+-x^-)}T_{--}(s) ~,
\eal
\eeq
where the location $x_0$ with coordinates $(x_0^+, x_0^-)$ is a reference point. Adjusting this point simply changes the constant $\phi_0$ in \eqref{eq:DilSol}.

Treating the matter sector quantum mechanically, the trace of the stress tensor no longer vanishes. Instead, for a metric written in conformal gauge,
\beq
\rmd s^2 = -e^{2\omega(x^+,x^-)}\rmd x^+\rmd x^- ~,
\eeq
the conformal anomaly leads to a non-zero off-diagonal component on curved backgrounds:
\beq \label{eq:StressOffDiag}
\langle T_{+-}\rangle = - \frac c{12\pi}\partial_+\partial_-\omega ~.
\eeq
Using the conformal anomaly as input, the diagonal components can be found by integrating the continuity equation $\nabla_\mu T^{\mu\nu} = 0$ \cite{Christensen:1977jc}
\beq \label{eq:StressDiag}
\langle T_{\pm\pm}\rangle = -\frac c{12\pi}\left(\partial_{\pm}^2\omega-(\partial_\pm\omega)^2\right) + :T_{\pm\pm}:~.
\eeq
The left-hand side is the covariant stress tensor that appears on the right-hand side of the semi-classical equations of motion and the second term on the right denotes the normal-ordered stress tensor.\footnote{Strictly speaking, normal ordering is only defined for free theories which is the case considered in this paper. More generally, one could replace $:T_{\pm\pm}:$ by a function $t_\pm(x^\pm)$ that parametrizes the flux seen by a local observer \cite{Fabbri:2005mw}.} This quantity depends on the choice of state and corresponds to left and right-moving radiation as measured by a local observer. 

Using the AdS metric \eqref{eq:AdSmetric} we see that 
\beq
\bal
\langle T_{\pm\pm}\rangle &= :T_{\pm\pm}: ~,\\
\langle T_{\pm\mp}\rangle &= \frac c {12\pi(x^+-x^-)^2} ~.
\eal
\eeq
Taking into account the conformal anomaly the dilaton solution is just modified by a constant:
\beq
\phi = \phi_0 -\frac{2\phi_r}{x^+-x^-} + 8\pi G_2\left(I_+ - I_- +  \frac{c}{24\pi} \right)~.
\eeq

\subsubsection*{Extremal Black Hole Solution}
We now couple this JT gravity solution to a non-gravitational Minkowski region described by metric $\rmd s^2 = -\rmd x^+\rmd x^-$. For simplicity, we consider the CFT to be at zero temperature, which sets $:T_{\pm\pm}: = 0$ and is appropriate to describe an extremal black hole \cite{Almheiri:2019yqk}. We then consider an interval $R$ whose endpoints we denote with $[a_i,b_i]$, where $i=(1,2)$. In Kruskal coordinates a point denoted by $a_i$ or $b_i$ is shorthand for $x^\pm = a_i^\pm$ or $x^\pm = b_i^\pm$. When we are working at a $t=0$ slice, we write $x=a_i$ or $x=b_i$ to denote the endpoints and drop the $\pm$ superscript. Note that $a_1$ and $b_1$ are negative as they are located in the AdS$_2$ region.

\begin{figure}[t]
\centering
\begin{tikzpicture}[scale=.8]

\fill [gray!10] (-0.5,-4.5) rectangle (3,4.5) ;

\draw[very thick] (3,-4.5) -- (3,4.5) node[pos=0,below]{$x=0$} ;
\draw[very thick] (-0.5,-4.5) -- (-0.5,4.5);
\draw[thick] (3,3.5) -- (6.5,0);
\draw[thick] (6.5,0) -- (3,-3.5);
\draw[thick] (3,3.5) -- (-0.5,0);
\draw[thick] (-0.5,0) -- (3,-3.5);

\draw[thick,red] (.5,0) -- (1.25,.75) node[pos=0,left]{$a_1^\pm$};
\node[red] at (1.2,0) {$I$};
\draw[thick,red] (1.25,.75) -- (2,0);
\draw[thick,red] (.5,0) -- (1.25,-.75);
\draw[thick,red] (1.25,-.75) -- (2,0) node[pos=1,right]{$b_1^\pm$};

\draw[thick,red] (4,0) -- (4.75,.75) node[pos=0,left]{$a_2^\pm$};
\node[red] at (4.7,0) {$R$};
\draw[thick,red] (4.75,.75) -- (5.5,0);
\draw[thick,red] (4,0) -- (4.75,-.75);
\draw[thick,red] (4.75,-.75) -- (5.5,0) node[pos=1,right]{$b_2^\pm$};

\draw[->,thick] (7,2) -- (8,3) node[right]{$x^+$};
\draw[->,thick] (7,2) -- (6,3) node[left]{$x^-$};
\draw[->,thick] (10,2) -- (10,3) node[left]{$t$};
\draw[->,thick] (10,2) -- (11,2) node[below]{$x$};

\end{tikzpicture}
\caption{AdS$_2$ region (shaded gray) at $x\leq 0$ coupled to a flat non-gravitational bath region at $x>0$. Two intervals are located at $t=0$ with $x\in[a_1,b_1]$ and $x\in[a_2,b_2]$. The two diamonds indicate the domain of dependence of the two intervals.}
\label{fig:AdS2setup}
\end{figure}
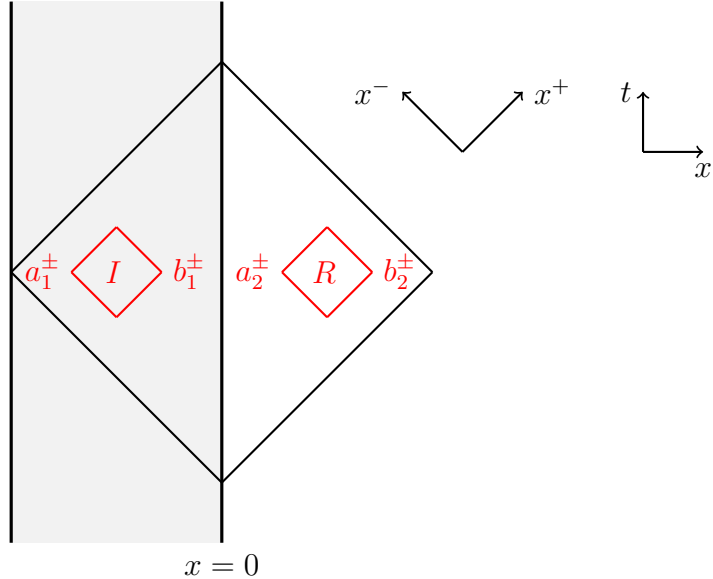

We now consider an interval $R$ at $t=0$ given by $x\in [a_2,b_2]$ that is entangled with the bulk gravity region. To compute the fine-grained entropy of $R$, we use the island formula \cite{Penington:2019npb,Almheiri:2019psf},
\beq
S(R) = \text{min, ext}_{a_1,b_1}\left(\frac{\text{Area}(\partial I)}{4G_2} + S_{\rm vN}(R\cup I)\right) ~.
\eeq
Here we allow for an island region $I$ located at $t=0$ and $x\in[a_1,b_1]$ whose endpoints we extremize. This setup is depicted in Figure \ref{fig:AdS2setup}.

The von Neumann entropy of $n$ disjoint intervals in a CFT consisting of free fermions in flat space was derived in \cite{Casini:2009vk}. In our case, we have two intervals; one is located in the flat bath region and the other in the AdS region. The resulting entropy can be obtained by a Weyl transformation of the flat space result. For two intervals $x=[a_1,b_1]$ and $x=[a_2,b_2]$ the flat space entropy is \cite{Casini:2009vk}\footnote{Note that (63) of \cite{Casini:2009vk} has a minor typo. The first sum on the second line should read: $\sum_{ij}\log|b_i-a_j|$.}
\beq
S_{\rm vN}^{\rm flat} = \frac c3\log\left[\frac{(b_1-a_1)(a_2-b_1)(b_2-a_1)(b_2-a_2)}{(a_2-a_1)(b_2-b_1)\epsilon^2}\right] ~.
\eeq
Transforming this result such that the interval $x=[a_1,b_1]$ is located on a background with non-trivial conformal factor we write the result in terms of the cross ratio
\beq
z = \frac{(b_2-a_2)(b_1-a_1)}{(b_2-b_1)(a_2-a_1)} ~,
\eeq
in two different ways:
\beq
\bal
S_{\rm vN}(R\cup I) &= \frac{c}{3}\log\left[\frac{(b_2-a_2)(b_1-a_1)}{\epsilon^2\exp(-\frac12[\omega(a_1)+\omega(b_1)])}\right] + \frac c3\log (1-z) ~, \\
&=\frac{c}{3}\log\left[\frac{(a_2-b_1)(b_2-a_1)}{\epsilon^2\exp(-\frac12[\omega(a_1)+\omega(b_1)])}\right] + \frac c3\log z ~. 
\eal
\eeq
The first line makes the limit $z\rightarrow 0$ easy to see, while the second does the same for the limit $z\rightarrow 1$. Here $e^{\omega(x)}$ is the conformal factor of the metric which in Poincaré coordinates in AdS$_2$ takes the form
\beq
e^{\omega(x)} = -\frac{\ell}{x} ~.
\eeq
This expression for the von Neumann entropy for two intervals is specific to the free fermion theory and only takes a universal form in particular limits.

One such limit is the OPE limit $z\to 1$ where the von Neumann entropy reduces to the sum of the individual entropies of the complementary regions. Alternatively, we can consider a limit of large central charge where the entropy takes a similar form depending whether $z\lessgtr \frac12$ \cite{Hartman:2013mia}:
\beq
S_{\rm vN}(R\cup I) =
\begin{cases}
\frac{c}{3}\log\left[\frac{(b_2-a_2)(b_1-a_1)}{\epsilon^2\exp(-\frac12[\omega(a_1)+\omega(b_1)])}\right] \quad (z<\frac12) \\
\frac{c}{3}\log\left[\frac{(a_2-b_1)(b_2-a_1)}{\epsilon^2\exp(-\frac12[\omega(a_1)+\omega(b_1)])}\right] \quad (z>\frac12)~.
\end{cases}
\eeq
Corrections to this large central charge result have been studied in \cite{Barrella:2013wja}. A non-trivial island only contributes for the case $z>1/2$ \cite{Rolph:2021nan}. Thus, working in this regime, the fine-grained entropy is given by
\beq
S(R) = \text{min,ext}_{a_1,b_1}\left(\frac{\phi(a_1)+\phi(b_1)}{4G_2} +  \frac{c}{3}\log\left[\frac{(a_2-b_1)(b_2-a_1)}{\epsilon^2\exp(-\frac12[\omega(a_1)+\omega(b_1)])}\right]\right) ~.
\eeq
Extremizing over $(a_1,b_1)$ we find that there is only one (consistent) solution in which the entropy is real and the island located in the bulk region.

The location of the quantum extremal surface expressed in terms of the length scale $L:=3\phi_r/(2cG_2)$ is
\beq
\bal \label{eq:IslandLocation}
a_1 &= -\frac12\left(b_2+L+\sqrt{b_2^2+6b_2L+L^2}\right) ~, \\
b_1 &= -\frac12\left(a_2+L+\sqrt{a_2^2+6a_2L+L^2}\right) ~.\\
\eal
\eeq
As usual in JT gravity, $L$ measures the deviation away from pure AdS$_2$.

Note that when the interval in the bulk region includes the `natural length' $L$, we can take the limit $a_2\ll L \ll b_2$, where the island simplifies to
\beq
a_1  = -b_2 ~, \quad b_1 = -L ~.
\eeq

In addition to the non-trivial QES there is always a trivial (vanishing) island. In that case, the fine-grained entropy is just given by $S_{\rm vN}(R)$. The fine-grained entropy is given by the minimum of these two contributions:
\beq
S(R) = \text{min}\left(S_{\rm vN}(R), S_{\rm QES}(R)\right) ~,
\eeq
with
\beq
\bal
S_{\rm vN}(R) &= \frac c3\log\left[\frac{b_2-a_2}{\epsilon}\right]~, \\
S_{\rm QES}(R) &= \frac{\phi_0+\phi_r/L}{2G_2} + \frac{\phi_r}{4G_2L}\log\left[\frac{4b_2\ell^2 L}{\epsilon^4}\right]~.
\eal
\eeq
One peculiar feature of this two-dimensional model of an extremal black hole is that its island lies \emph{outside} the black hole horizon \cite{Almheiri:2019yqk}. Consequently, the modular curvature we will study later also only probes the region outside the horizon. However, for a generic model of an evaporating black hole in JT gravity, such as originally studied in \cite{Penington:2019npb,Almheiri:2019psf}, the location of the island is behind an horizon. Applying our method to these (more complicated) models should yield (geometrical) information behind horizons.

Now that we have reviewed JT gravity including its islands, we will next construct the associated modular Hamiltonian. We will perform modular transport in two cases; both for a single interval and for the two-interval island case.

\subsection{Modular Transport for a Single Interval}
Similar to holography in higher dimensions, in JT gravity we can consider the bulk AdS$_2$ dual of the boundary modular Hamiltonian defined in the CFT$_1$. The relation between the bulk and boundary modular Hamiltonians is given by a two-dimensional version of the JLMS formula \cite{Jafferis:2015del}:
\beq
H_{\rm bdy} = \frac{\hat A}{4G_2} + H_{\rm bulk} + {\cal O}(G_2)~.
\eeq
Here $\hat A$ is the area operator and $H_{\rm bulk}$ is the bulk modular Hamiltonian which acts on the matter fields.

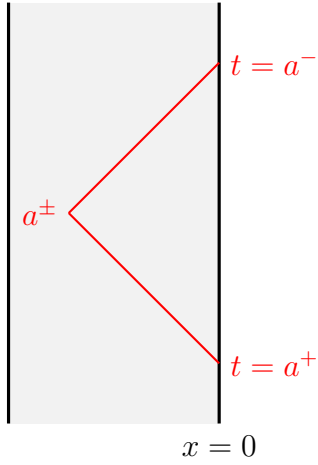
\begin{figure}[t]
\centering
\begin{tikzpicture}[scale=.8]

\fill [gray!10] (-0.5,-3.5) rectangle (3,3.5) ;

\draw[very thick] (3,-3.5) -- (3,3.5) node[pos=0,below]{$x=0$} ;
\draw[very thick] (-0.5,-3.5) -- (-0.5,3.5);

\draw[thick,red] (.5,0) -- (3,2.5) node[pos=0,left]{$a^\pm$} node[pos=1,right]{$t=a^-$};
\draw[thick,red] (.5,0) -- (3,-2.5) node[pos=1,right]{$t=a^+$};

\end{tikzpicture}
\caption{Two-dimensional analogue of an RT surface, which is a single point. From this point we shoot two lightrays to define a causal diamond. This defines a time interval on the AdS boundary.}
\label{fig:RTsurface}
\end{figure}

To get some intuition for this expression, we consider the two-dimensional analogue of an RT surface, which is a single point that we denote by $x^\pm = a^\pm$. From this point, we shoot lightrays to the boundary to form a causal diamond, as in Figure \ref{fig:RTsurface}.
Taking the state of the two-dimensional bulk matter to be the vacuum, i.e. $:T_{\pm\pm}: = 0$, the bulk modular Hamiltonian acts as the boost Killing vector that leaves the diamond invariant. We note that in this section we do not impose that the interval defining the diamond lies on a $t=0$ slice. The vector field preserving the diamond is given by
\beq \label{eq:KV-bdy}
\zeta^\mu\partial_\mu = \frac{(a^+-x^+)(a^--x^+)}{a^+-a^-}\partial_+ + \frac{(a^+-x^-)(a^--x^-)}{a^+-a^-}\partial_- ~.
\eeq
Thus, using \eqref{eq:localModHam}, the bulk modular Hamiltonian takes the form
\beq \label{eq:matterModHam}
H_{\rm bulk} = 2\pi \left( \int^{\frac12(a^++a^-)}_{a^+} \rmd s \frac{(a^+-s)(a^--s)}{a^+-a^-}T_{++}(s) - \int^{\frac12(a^++a^-)}_{a^-} \rmd s \frac{(a^+-s)(a^--s)}{a^+-a^-}T_{--}(s) \right) ~,
\eeq
which we recognize as the part of the dilaton solution coupled to the CFT$_2$ evaluated at $(x^+,x^-)=(a^+,a^-)$. We can now make the following identification \cite{Callebaut:2018nlq, Callebaut:2018xfu}:
\beq
H_{\rm bdy} = \frac{\hat \phi}{4G_2} ~,
\eeq
which relates
\beq
\bal
\langle \hat A \rangle &= \phi_0 -\frac{2\phi_r}{a^+-a^-} ~, \\
\langle H_{\rm bulk} \rangle &=  2\pi\left(I_+ - I_-\right) ~.
\eal
\eeq
Here we dropped the constant piece given by the central charge, which will drop out of the generators of modular transport. The area term is the Noether charge associated with the Killing vector that preserves the diamond, evaluated at the bulk point (see e.g. Appendix C of \cite{Pedraza:2021cvx}). Thus, just as in higher-dimensional cases, boundary modular flow equals bulk modular flow.

We are now interested in studying the behavior of the modular Hamiltonian in the presence of an island. As before, we couple the AdS$_2$ bulk to a non-gravitating bath region, see Figure \ref{fig:AdS2setup}. When we consider an interval in the bath in the presence of an island, the boundary modular Hamiltonian is given by an expression similar to the island formula for the fine-grained entropy \cite{Chen:2019iro}:
\beq \label{eq:IslandModHam}
H_{\rm bdy} = \frac{\hat A(\partial I)}{4G_2} + H_{\rm bulk}(A\cup I) ~.
\eeq
A remarkable property this island generalization of the JLMS formula is that the left-hand side is proposed to be the exact fine-grained modular Hamiltonian associated to the quantum state of the boundary theory, but the right-hand side is an expression evaluated in the semi-classical gravity theory. Apparently, semi-classical gravity has access to some fine-grained information of the full theory. Accordingly, we can evaluate the modular Hamiltonian, just by knowing its form in the semi-classical theory.

Evaluated on a trivial island ($I=\varnothing$), \eqref{eq:IslandModHam} reduces to the matter modular Hamiltonian in the non-gravitational bath. In the presence of a non-trivial island, we instead have to understand the modular Hamiltonian of two disjoint intervals. In general, this two-interval modular Hamiltonian is a complicated non-local expression whose closed form is only known in special cases. In Sec. \ref{sec:IslandCurvature} we therefore consider a matter sector which consists of free massless fermions, for which the modular Hamiltonian has been derived in \cite{Casini:2009vk}. Before doing so, we first consider a single interval in the bulk and explicitly show how (bulk) modular transport probes the Riemann curvature.

\subsubsection*{Modular Curvature of a Bulk Interval}
We now fix an interval in the bulk that defines a causal diamond. We then consider the modular Hamiltonian associated to this diamond and show how its modular curvature is related to the Riemann curvature. We stress that this is a toy example, because we define the bulk diamond just by fixing its endpoints in a gravitational theory. This is not diffeomorphism invariant and the resulting expressions might therefore be gauge dependent. Still, this will be an instructive example because the resulting expressions will naturally translate to the situation where we consider an island which we do define in a diff-invariant manner by specifying an interval in the bath region. 

\begin{figure}[t]
\centering
\begin{tikzpicture}[scale=.8]

\fill [gray!10] (-0.5,-2) rectangle (3.5,2) ;

\draw[very thick] (3.5,-2) -- (3.5,2) node[pos=0,below]{$x=0$} ;
\draw[very thick] (-0.5,-2) -- (-0.5,2);

\draw[thick,red] (.5,0) -- (1.5,1) node[pos=0,left]{$a^\pm$};
\draw[thick,red] (1.5,1) -- (2.5,0);
\draw[thick,red] (.5,0) -- (1.5,-1);
\draw[thick,red] (1.5,-1) -- (2.5,0) node[pos=1,right]{$b^\pm$};

\end{tikzpicture}
\caption{Causal diamond in the bulk associated to the interval $x^\pm = [a^\pm,b^\pm]$.}
\label{fig:DiamondSetup}
\end{figure}
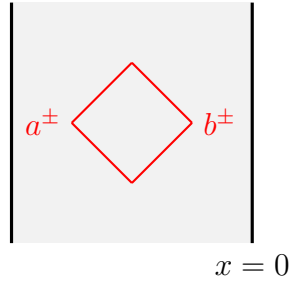

Keeping in mind this subtlety, we now show how the bulk Riemann curvature can be obtained from modular transport when we consider a causal diamond in the bulk. The endpoints of the interval specifying the causal diamond are given by $x^\pm=[a^\pm,b^\pm]$, see Figure \ref{fig:DiamondSetup}.
In vacuum, the matter modular Hamiltonian is given by the expression \eqref{eq:localModHam} where $\zeta^\mu$ is the conformal Killing vector that preserves the diamond, i.e.
\beq \label{eq:CKV}
\zeta^\mu\partial_\mu = \frac{(a^+-x^+)(b^+-x^+)}{a^+-b^+}\partial_+ - \frac{(a^--x^-)(b^--x^-)}{a^--b^-}\partial_- ~. 
\eeq
We note that the KV \eqref{eq:KV-bdy} that preserves the half diamond attached to the boundary is a special case of this general CKV evaluated at $b^\pm = a^\mp$.
\eqref{eq:CKV} is the conformal Killing vector for any two-dimensional metric written in conformal gauge:
\beq
\rmd s^2 = -e^{2\omega(x^+,x^-)}\rmd x^+\rmd x^- ~.
\eeq

We will now show how the bulk modular Hamiltonian can be used to probe the curvature. As before, we choose to represent the modular Hamiltonian directly by the conformal Killing vector, i.e.
\beq
K = 2\pi \zeta^\mu\partial_\mu ~.
\eeq
We split this vector into a left-moving and right-moving piece:
\beq
K = K_{(+)} + K_{(-)} ~.
\eeq
Focusing on the $+$ sector the generators of modular transport are now given by
\beq
\bal
V_{a^+} &= +\frac1{2\pi}\partial_{a^+}K ~, \\
V_{b^+} &= -\frac1{2\pi}\partial_{b^+}K ~,
\eal
\eeq
with similar expressions for $V_{a^-}, V_{b^-}$. They obey the expected commutation relations of modular scrambling modes \cite{DeBoer:2019kdj},
\beq
[V_{a^+},K] =+2\pi\, V_{a^+} ~, \qquad [V_{b^+},K] =-2\pi\, V_{b^+} ~.
\eeq
The modular curvature is now computed to be
\beq \label{eq:ModCurvSingle}
{\cal R}_{a^+b^+} := [V_{a^+},V_{b^+}] = \frac{2}{(a^+-b^+)^2}\left(\frac{K_{(+)}}{2\pi}\right) ~.
\eeq

To relate the modular curvature to the Riemann curvature, we can use \eqref{eq:ConformalCurvature}. Because \eqref{eq:CKV} is the CKV of a causal diamond in any two-dimensional metric, this gives a result that is valid in general. As before, we express the change in the modular Hamiltonian in terms of $K'_{a^+}$, $K'_{b^+}$ defined as:
\beq \label{eq:Kprime}
K'_{a^+} = \frac1{(2\pi)^2}[V_{a^+},K] ~, \quad K'_{b^+} = \frac1{(2\pi)^2}[V_{b^+},K] ~.
\eeq
Evaluating the right-hand side of \eqref{eq:ConformalCurvature} we find that this expression simplifies and reduces to
\beq \label{eq:ConformalCurvature2D}
{\cal R}^+_{a^+b^+} = V_{a^+}^+R_{-++}^{~~~~~+}K'{}_{b^+}^+ K^- -\nabla_+V_{a^+}^+V_{b^+}^+ + \frac12\Omega^+~.
\eeq
This equation relates the modular curvature of a single interval in any two-dimensional spacetime to its Riemann curvature.

We will now show how the presence of an island modifies the modular curvature.

\subsection{Modular Transport in the Presence of an Island} \label{sec:IslandCurvature}
 We will now study modular transport in the presence of a non-trivial quantum extremal surface. Before doing so, let us first reiterate the general idea. We consider a gravitating two-dimensional AdS space glued to a non-gravitating flat `bath' region. Using modular transport in the bath region we want to understand if we can extract the bulk Riemann curvature associated to an island region, using the JLMS relation including the possibility of an island \eqref{eq:IslandModHam}.
 
 Similar to the island formula that computes the fine-grained entropy of Hawking radiation, \eqref{eq:IslandModHam} states that the \emph{exact} modular Hamiltonian of a subregion of the bath (which is unknown in general) equals a simpler semi-classical expression involving the semi-classical bulk modular Hamiltonian of two disjoint intervals. For generic matter, the modular Hamiltonian of disjoint intervals is unknown. However, for free fermion theories, \cite{Casini:2009vk} derived an explicit formula. Consequently, we will now restrict to this case.

\subsubsection*{Modular Hamiltonian of Disjoint Intervals} 
The modular Hamiltonian of free fermions on $n$ disjoint intervals in $1+1$-dimensional flat space was derived in \cite{Casini:2009vk}. This model has been studied in the context of quantum extremal surfaces in \cite{Chen:2019iro} and modular transport in \cite{Chen:2022nwf}. The latter reference makes use of an isomorphism---referred to as a multi-local symmetry---to transform the complicated non-local expression of the modular Hamiltonian of two disjoint intervals into a tractable local expression for two fermions on a half line where modular flow acts as a boost.

\begin{figure}[t]
\centering
\begin{tikzpicture}[scale=.8]

\draw[->,thick] (-1,0) -- (3,4) node[right]{$x^+$};
\draw[->,thick] (-1,0) -- (3,-4) node[right]{$-x^-$};

\draw[dashed] (0,0) -- (-.5,.5) node[pos=1,left]{$a_1^+$};
\draw[dashed] (0,0) -- (-.5,-.5) node[pos=1,left]{$a_1^-$};
\draw[dashed] (2,0) -- (0.5,1.5) node[pos=1,left]{$b_1^+$};
\draw[dashed] (2,0) -- (0.5,-1.5) node[pos=1,left]{$b_1^-$}; 

\draw[dashed] (4,0) -- (1.5,2.5) node[pos=1,left]{$a_2^+$};
\draw[dashed] (6,0) -- (2.5,3.5) node[pos=1,left]{$b_2^+$};
\draw[dashed] (4,0) -- (1.5,-2.5) node[pos=1,left]{$a_2^-$};
\draw[dashed] (6,0) -- (2.5,-3.5) node[pos=1,left]{$b_2^-$};

\draw[|-|,thick,red] (0,0) -- (2,0) ;
\draw[|-|,thick,red] (4,0) -- (6,0) ;

\end{tikzpicture}
\caption{Two disjoint intervals whose endpoints are labeled by lightcone coordinates.}
\label{fig:TwoIntervals}
\end{figure}
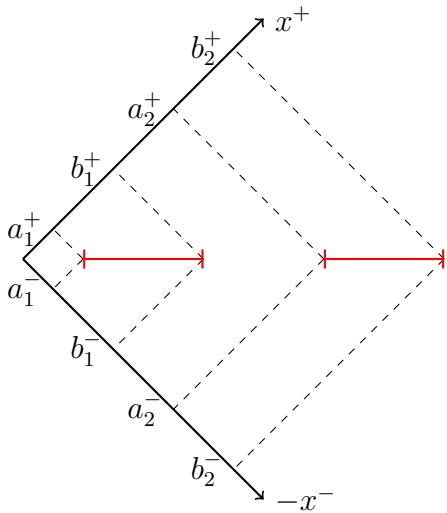

We focus on two intervals and label the endpoints as indicated in Figure \ref{fig:TwoIntervals}.
We first consider the non-gravitational case where these two intervals are located on flat space. As we will argue, the generalization to AdS$_2$ is straightforward; for our purpose of studying modular flow, it is actually essentially unmodified from the flat space case. The modular Hamiltonian splits into two pieces: $H=H_{(+)}+H_{(-)}$. In two dimensions the left and right-moving sector completely decouple so it suffices to just consider one sector. We focus on the $x^+$ sector, and instead of keeping track of the superscript, we just write $x$. We then have \cite{Casini:2009vk,Chen:2022nwf}
\beq \label{eq:ModHamDisjoint}
H_{(+)} = 2\pi \int \rmd x\left(h_T(x)T(x) + h_C(x)\psi^\dagger(x)\psi(\tilde x)\right) ~.
\eeq
The coefficients of the operators and the map $\tilde x(x)$ are
\beq
\bal \label{eq:ModHamCoefficients}
h_T(x) &= \left(\partial_xy(x)\right)^{-1}  ~, \\
h_C(x) &= \frac{\partial_x \tilde x}{x-\tilde x}h_T(x) ~, \\
y(x) &= \log\left[-\frac{(x-a_1)(x-a_2)}{(x-b_1)(x-b_2)}\right] ~, \\
\tilde x(x) &= \frac{a_1 a_2(x-b_1-b_2) - b_1 b_2(x-a_1-a_2)}{(a_1 + a_2 -b_1 - b_2)x - a_1 a_2 + b_1 b_2} ~,
\eal
\eeq
and the fermion stress tensor is given by
\beq
T(x) = \frac12\left(\partial_x \psi^\dagger\psi-\psi^\dagger\partial_x \psi\right)~.
\eeq

Since we are focusing on the + sector, the coefficients $a_i,b_i$ refer to the + components. The map $\tilde x(x)$ is quasi-local in the sense that it exchanges points in the two intervals. The second term in the modular Hamiltonian therefore couples fermions non-locally, but point wise. This fact was used in \cite{Chen:2019iro} to understand how an operator in an island region can be mapped to a bath region under modular flow.\footnote{The reference \cite{Chen:2019iro} uses different conventions to label the intervals. The relation between our and their conventions is given by $(a_{1,2}^+,b_{1,2}^+)_{\rm us}\to (a_{1,2}^+,b_{1,2}^+)_{\rm them}$ and $(a_{1,2}^-,b_{1,2}^-)_{\rm us}\to (b_{2,1}^-,a_{2,1}^-)_{\rm them}$. Thus, for our focus on the + sector, the two conventions coincide.}

To highlight the difference between the local modular Hamiltonian for a single interval and its non-locality for disjoint intervals, we display the evolution of the local fermion operators under modular flow. For this purpose it is useful to consider the function $y(x)$ defined in \eqref{eq:ModHamCoefficients}. This function runs from $[-\infty,+\infty]$ in each interval $[a_1,b_1]$ and $[a_2,b_2]$. A value of $y(x)$ corresponds to one point in each interval, which we denote by $x_i(y)$. We can then define a fermion operator in each interval as
\beq
\tilde \psi_i(y) := \sqrt{\frac{dx_i}{dy}}\psi(x_i(y)) ~.
\eeq
Under modular flow, the positions of the operators evolve with the modular time parameter $s$ as $y(s) = y(0) + 2\pi s$ and the operators transform as \cite{Casini:2009vk}
\beq
\psi_i(y(s)) = R_i^{\,\,\,j}\psi_j(y(0)) ~.
\eeq
Here $R$ is a rotation matrix
\beq
R_i^{\,\,\,j} =
\begin{pmatrix}
\cos \theta(s) & \quad-\sin\theta(s) \\
\sin\theta(s) & \quad\cos\theta(s)
\end{pmatrix}
~,
\eeq
with angle
\beq
\theta(s) = \left[\arctan\left(\frac{(b_1+b_2-a_1-a_2)x_1(s)+(a_1a_2-b_1b_2)}{\sqrt{(b_1-a_1)(a_2-b_1)(b_2-a_1)(b_2-a_2)}}\right) - (s = 0 )\right] ~.
\eeq
Specifying an initial condition $y(0)$, this relation can be inverted to obtain the position of the operators under modular flow.

\begin{figure}[t]
\centering
\includegraphics[scale=.3]{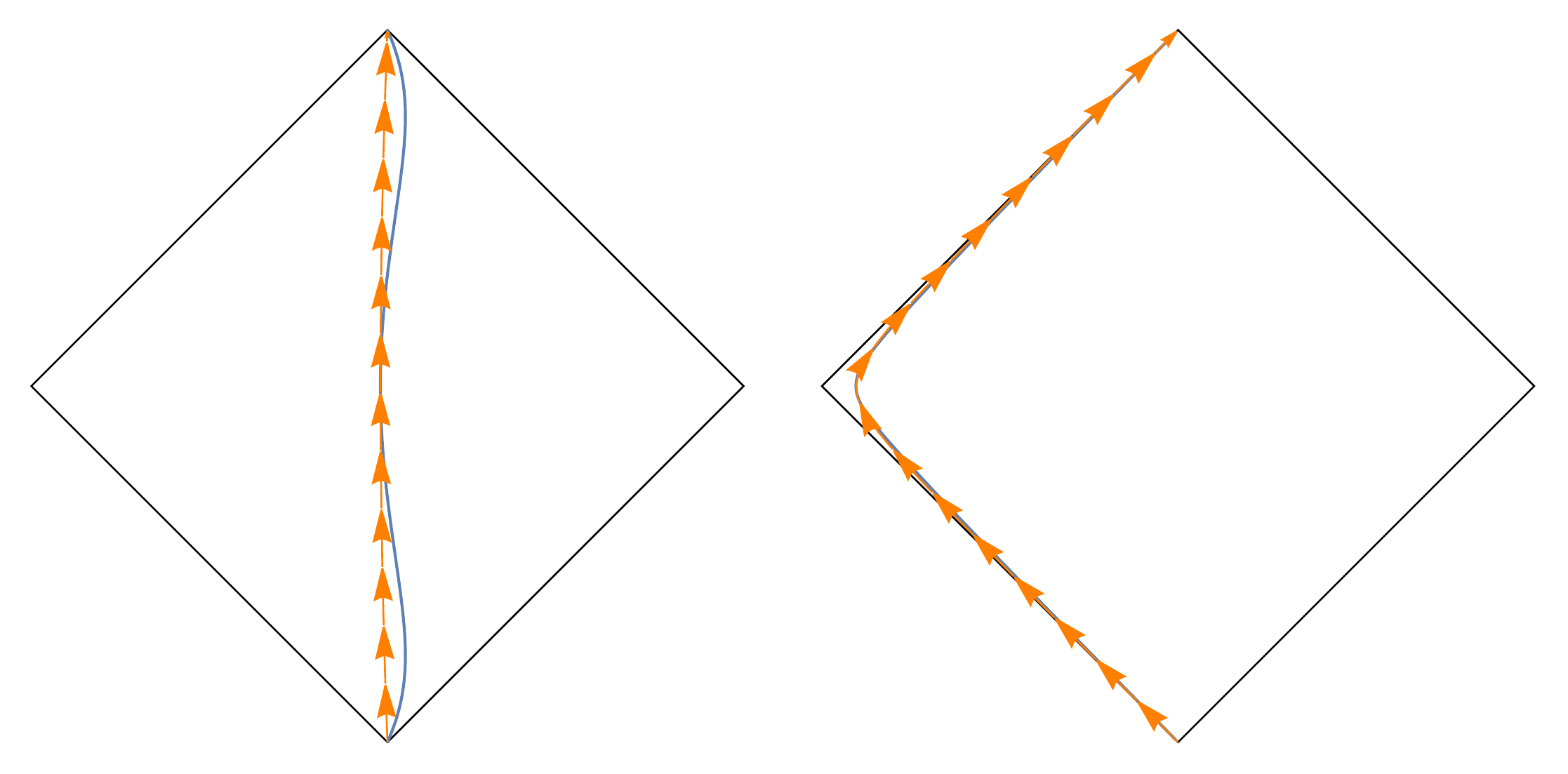}
\caption{Behavior of the position of the operators $\tilde\psi_i(y)$ under modular flow. The blue line corresponds to the modular flow trajectory and the orange trajectory corresponds to the CKV of each diamond. We see that, in general, modular flow in the bulk of the diamond does not coincide with the flow of the CKV of the diamond. Close to the edge of the diamond, however, they do coincide. We took $[a_1,b_1]=[-0.01,0.99]$, $[a_2,b_2]=[1.1,2.1]$ and $y(0) = -1$.}
\label{fig:DoubleFlow1}
\end{figure}

\begin{figure}[t]
\centering
\includegraphics[scale=.3]{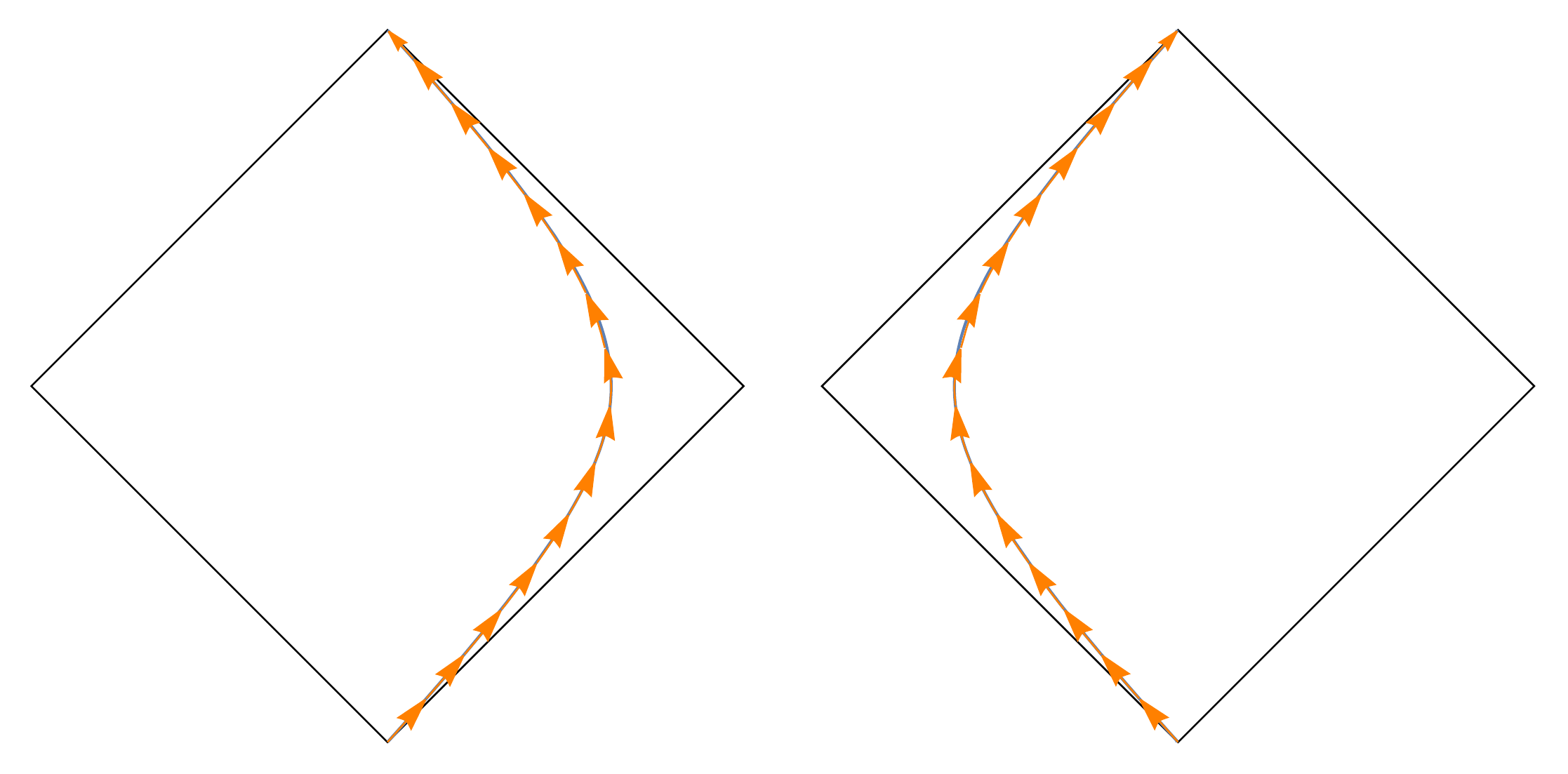}
\caption{Behavior of the position of the operators $\tilde\psi_i(y)$ under modular flow. The blue line corresponds to the modular flow trajectory and the orange trajectory corresponds to the CKV of each diamond. If we consider operators close to the edge of the diamond, the modular flow trajectory coincides with the CKV of the diamond. We took $[a_1,b_1]=[-0.01,0.99]$, $[a_2,b_2]=[1.1,2.1]$ and $y(0) = 0$.}
\label{fig:DoubleFlow2}
\end{figure}

\begin{figure}[t]
\centering
\includegraphics[scale=.42]{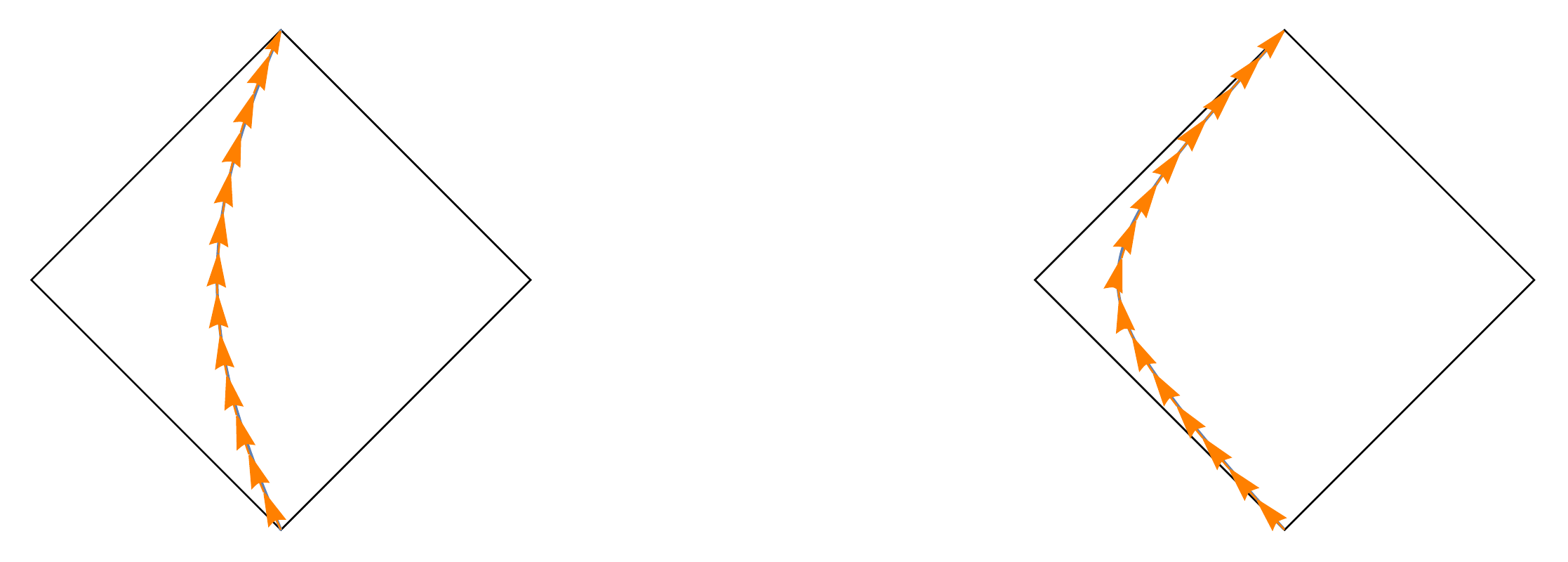}
\caption{Behavior of the position of the operators $\tilde\psi_i(y)$ under modular flow. We note that modular flow couples points at different values $y(x)$ in each interval. The blue line corresponds to the modular flow trajectory and the orange trajectory corresponds to the CKV of each diamond. When the diamonds are far away from each other the modular flow trajectory coincides with the CKV of the diamond. We took $[a_1,b_1]=[-0.01,0.99]$, $[a_2,b_2]=[2,3]$ and $y(0) = -1$.}
\label{fig:DoubleFlow3}
\end{figure}

Fixing the length of both diamonds to be equal, we demonstrate the behavior of modular flow in three illustrative examples in Figures \ref{fig:DoubleFlow1}, \ref{fig:DoubleFlow2} and \ref{fig:DoubleFlow3}. %
We notice that for diamonds that are far away from other, modular flow follows the trajectory of the CKV of the diamond. Indeed, this is also easy to see analytically. In the limit where the cross-ratio $z\to 0$, the non-local piece of the modular Hamiltonian vanishes. One way of taking this limit is by shrinking one of the intervals to zero size, let's say $a_2 \to b_2$. The modular Hamiltonian then becomes the CKV of the remaining interval:
\beq
\bal
\lim_{a_2 \to b_2} h_T(x) &= \frac{(a_1-x)(b_1-x)}{a_1-b_1} ~, \\
\lim_{a_2 \to b_2} h_C(x) &= 0 ~.
\eal
\eeq

In addition, the figures also make it clear that modular flow follows the trajectory of a CKV when we consider an operator that is close to the edge of the diamond. This behavior seems similar to the fact that the modular Hamiltonian of excited states reduces to the boost Killing vector close to the boundary, although here there still is a non-local piece. Thus, the operators $\tilde \psi_1(y)$ and $\tilde \psi_2(y)$ still mix but their trajectories are described by the CKV of their respective diamonds.

In principal, one can obtain the generators of modular transport by varying \eqref{eq:ModHamDisjoint}. However, away from a simplifying limit where the non-local piece vanishes, this procedure quickly becomes unwieldy when studying modular transport. For this reason, in cases where the non-local contribution cannot be dropped, it is more convenient to make use of the multi-local symmetry utilized in \cite{Chen:2022nwf} to obtain a more tractable expression for the modular Hamiltonian that is local. The map to a local modular Hamiltonian involves an isomorphism $\beta$ that maps two species of fermion $\Psi_{i=1,2}$ on a half-line $I=[0,\infty]$ to a single fermion $\psi$ on two disjoint intervals $I_1\cup I_2=[a_1,b_1]\cup [a_2,b_2]$. This map involves a coordinate transformation to  `$X$-space':\footnote{We refer to the representation \eqref{eq:ModHamDisjoint} as $x$-space, which is the same as the $z$-space in \cite{Chen:2022nwf}.}
\beq
X(x) = -\frac{(x-a_1)(x-a_2)}{(x-b_1)(x-b_2)} ~.
\eeq
It has two inverses $x_{1,2}(X)$, which send the half-line $[0,\infty]$ to either the interval $[a_1,b_1]$ or $[a_2,b_2]$. For more details regarding this multi-local symmetry we refer the reader to \cite{Wong:2018svs,Chen:2022nwf}. Here we just note that the map $\beta$ acts on fermions in $X$-space as:
\beq
\bal
\beta[\Psi_i(X)] = \sum_{j=1,2} U_{ij}\sqrt{\partial_X x_j(X)}\psi(x_j(X)) ~, \\
\beta[\Psi^\dagger_i(X)] = \sum_{j=1,2} U_{ij}\sqrt{\partial_X x_j(X)}\psi^\dagger(x_j(X)) ~,
\eal
\eeq
where $U=e^{\epsilon \Theta(X)}$ is an orthogonal matrix. The parameter $\Theta(X)$ is defined as
\beq
\Theta(X) = \int^X\rmd Y \frac{\sqrt{\partial_Yx_1(Y)\partial_Yx_2(Y)}}{x_1(Y)-x_2(Y)} ~,
\eeq
and implements a gauge transformation that introduces a monodromy around the branch points of the map $X(x)$.

To explicitly evaluate the action of the map, it will be useful to expand $U_{ij} = \delta_{ij} + \epsilon_{ij}\Theta(x) + \dots$, where $\epsilon_{ij}$ is the Levi-Civita symbol. The inverse map $\beta^{-1}$ acts on $H_{(+)}$ as
\beq \label{eq:ModHamX}
\beta^{-1} \circ H_{(+)} = 2\pi \int\rmd X X\left(T_{11}(X) + T_{22}(X)\right) ~.
\eeq
We see that in $X$-space the modular Hamiltonian simply acts as a boost on the two fermions defined in Rindler space. The stress tensors of the two fermions are given by
\beq
T_{ij}(X) = \frac12\left(\partial_X\Psi_i^\dagger(X)\Psi_j(X) - \Psi_i^\dagger(X)\partial_X\Psi_j(X)\right) ~.
\eeq
Varying \eqref{eq:ModHamX} gives a more tractable way of obtaining the generators of modular transport.

\subsubsection*{Modular Transport with Islands}
The expressions for the modular Hamiltonian and modular transport that we just presented are valid on flat space. However, our setup of interest consists of one (bath) interval located on flat space and one on AdS$_2$. Nonetheless, because we consider the fermion in the vacuum state, i.e. $:T_{\pm\pm}:\,=0$, we can see from \eqref{eq:StressOffDiag} and \eqref{eq:StressDiag} that the only difference between the stress tensor on flat space and AdS$_2$ is the conformal anomaly. This contribution can be safely dropped to study the dynamics of the modular Hamiltonian, such that we can readily apply the expression for the modular Hamiltonian on flat space to our situation of interest.

Now we will take some interesting limits. We first evaluate the modular Hamiltonian on the location of the island, given by \eqref{eq:IslandLocation}.  Working in the limit $a_2/L \ll 1 $ and $ b_2/L\gg 1$, keeping $L$ fixed, the location of the island is simply given by $(a_1,b_1)=(-b_2,-L)$. Since we are working with a CFT in the vacuum, we can equivalently describe this situation in terms of the complement of the two intervals which, in this limit, is just a single interval. We then expect the modular Hamiltonian to reduce to the single interval result, i.e. it should implement a boost with respect to the CKV of the causal diamond defined by the interval $x=[-L,0]$. Indeed, taking this limit we obtain
\beq
h_T(x)_{\rm QES} = \frac{x(L + x)}{L} ~,
\eeq
which corresponds to minus the $+$ component of the CKV of the diamond associated to the interval $x=[-L,0]$. As anticipated, this interval is the complement of the $[a_1,b_1]\cup [a_2,b_2]$ after imposing the QES condition and taking the limit $a_2/L\ll 1 $ and $ b_2/L \gg 1$. 

Because in this limit the modular Hamiltonian reduces to the CKV of the diamond $x=[-L,0]$, we expect to be able to still extract the Riemann curvature using modular transport. This expectation is essentially true, up to a subtlety involving the non-local contribution. Taking the same limit in the non-local coefficient, we obtain
\beq
h_C(x)_{\rm QES} =  \frac{2x(L+x)}{L(L+2x)}~.
\eeq
Perhaps surprisingly, this term is non-zero. One should be careful however in interpreting this result, as the map $\tilde x(x)$ becomes ill-defined in this limit as one interval has reduced to a point. Thus, the non-local coupling $\psi^\dagger(x)\psi(\tilde x)$ needs to be reinterpreted. Said differently, considering the modular Hamiltonian of a single interval and taking the single-interval limit of the two-interval modular Hamiltonian do not commute. Later, we will show that in $X$-space the single interval limit is regular and non-local contributions to modular transport vanish. For now, we will therefore simply ignore the non-local contribution.

After varying the modular Hamiltonian \eqref{eq:ModHamDisjoint}, \cite{Chen:2022nwf} obtained the generators of modular transport. As mentioned, for now we only keep the local terms.\footnote{This amounts to only keeping the first term proportional to $T(x)$ in \eqref{eq:ExplicitGenerators}.} The complete generators (including non-local terms) are displayed in Appendix \ref{sec:generators}.

As in the AdS$_3$/CFT$_2$ case, we define a left-moving and right-moving piece of the (local) modular Hamiltonian as $K = K_{(+)} + K_{(-)} $ with
\beq
K_{(\pm)} = \pm 2\pi\, h_T(x^\pm)\partial_\pm ~.
\eeq
In the single interval limit we are considering the non-zero generators are (where we dropped the $+$ superscript on $b_1,a_2$)
\beq
\bal
V_{b_1} &= -\frac1{2\pi}\partial_{b_1} K_{(+)} ~, \\
V_{a_2} &= +\frac1{2\pi}\partial_{a_2} K_{(+)} ~.
\eal
\eeq
We now obtain the expected commutation relations
\beq
[V_{b_1},K] = \partial_{b_1}K ~, \qquad [V_{a_2},K] = \partial_{a_2}K ~.
\eeq
The only non-zero component of the modular curvature is given by
\beq\label{eq:IslandCurv}
{\cal R}_{a_2b_1} = \frac2{L^2} h_T(x) ~.
\eeq
We recognize this expression as $2/(2\pi L^2)$ times the modular Hamiltonian. The coefficient is indeed $2/(b_1-a_2)^2$ evaluated on the island, as expected. Using the relation \eqref{eq:ConformalCurvature2D} we can now relate this modular curvature to the Riemann curvature.

Due to the generalization of the JLMS relation to include a possible island (see \eqref{eq:IslandModHam}), we see that it is possible to reconstruct the Riemann curvature far away from the bath region by acting with the exact generator of modular transport. Concretely, let us define the generator $V_{a_2}^Q$ in the exact theory. This generator is a complicated operator that (non-perturbatively) has information about the bulk. To distinguish the modular curvature computed using the generators $V^Q_{a_2}$ and $V^Q_{b_2}$ from the modular curvature in the semi-classical theory we refer to any commutator of an exact generator with another generator as the Quantum Extremal Modular Curvature (QEMC). Knowledge of this operator allows an observer in the boundary theory to extract geometrical information of islands in the bulk using the QEMC. Concretely, assuming the presence of an island in the semi-classical theory, $V_{a_2}^Q$ generates modular transport on the two disjoint intervals:
\beq
V_{a_2}^Q = V_{a_2} + \left(\frac{\partial b_1}{\partial a_2}\right)V_{b_1} ~.
\eeq
Accordingly, in the semi-classical theory, changing the endpoint of the bath interval also changes the location of the island via the QES condition. In the limit where the semi-classical generators are local, an observer in the exact theory can now obtain the Riemann curvature by computing
\beq \label{eq:x-spaceCurv}
[V_{a_2},V_{a_2}^Q] = \left(\frac{\partial b_1}{\partial a_2}\right)[V_{a_2},V_{b_1}] = (-2){\cal R}_{a_2b_1} ~.
\eeq
Here ${\cal R}_{a_2b_1}$ is the modular curvature in the semi-classical theory. The factor of $(-2)$ comes from the Jacobian.

This commutator might look a bit funny, since it involves the same endpoint $a_2$. However, other commutators involving one exact generator just reduce to their semi-classical values, i.e. $[V_{a_2},V_{b_2}^Q] = [V_{a_2},V_{b_2}]$ and $[V_{b_2},V_{a_2}^Q] = [V_{b_2},V_{a_2}]$, and don't probe the bulk region. Using the relation between the (local) modular curvature and Riemann curvature, i.e. \eqref{eq:ConformalCurvature2D}, we now see how the QEMC can be used to extract the Riemann curvature from modular flow. So far, we've worked in the limit where the generators of modular transport become local and saw how we could extract the Riemann curvature. Next, we go beyond this limit and compute the leading corrections to the geometric modular flow.

\subsubsection*{Beyond Geometric Flow}
To compute the leading corrections to the geometric limit of the QEMC considered so far, we will make use of the multi-local symmetry we described before. This symmetry allows for tractable expressions of the generators of modular transport. To compute the leading correction to the geometric expression for the modular curvature, we still consider the same commutator in the presence of an island, i.e.
\beq
[V_{a_2},V_{a_2}^Q] = \left(\frac{\partial b_1}{\partial a_2}\right)[V_{a_2},V_{b_1}]=\left(\frac{\partial b_1}{\partial a_2}\right){\cal R}_{a_2b_1}  ~,
\eeq
but now we consider the generators in $X$-space. The explicit expressions are given in Appendix \ref{sec:generators}. We then expand to leading order in an expansion of $a_2 / L \ll 1$ and $b_2 / L \gg 1$. Writing the commutator of interest in terms of the different zero modes that contribute we have:
\beq \label{eq:X-spaceCurv}
[V_{a_2},V^Q_{a_2}] = -\frac4{L^2}\left[\,Q_{22}^{(1)} + \sqrt{\frac{L}{8b_2}} \left(Q_{21}^{(0)}-Q_{12}^{(0)} + 2\left(Q_{12}^{(1)}+Q_{21}^{(1)} - Q_{21}^{(2)}\right) \right) \right] ~.
\eeq
The different zero modes are given by
\beq
\bal
Q_{ij}^{(0)} &= \int \rmd X \Psi_i(X)\Psi_j(X) ~, \\
Q_{ij}^{(1)} &= \int \rmd X X T_{ij}(X) ~, \\
Q_{ij}^{(2)} &= \int \rmd X X^2 \left(\partial_X\Psi^\dagger_i(X)\partial_X\Psi_j(X) - (i\leftrightarrow j)\right) ~.
\eal
\eeq
The first term in the expansion is proportional to just $Q_{22}^{(1)}$ and does not involve the fermion $\Psi_1$. As promised, we now see that---when transformed back to a \emph{single} interval in $x$-space---the first term also has a local interpretation as a boost implemented by a CKV. Explicitly, when acting with the map $\beta$ on the zero mode $Q_{22}^{(1)}$ to transform back to $x$-space, the term involving a derivative of the gauge parameter $\Theta(X)$ becomes proportional to the term $\psi^\dagger(x)\psi(\tilde x)$. We find that $\Theta'(X)$ vanishes in the limit $a_2 / L \ll 1$ and $b_2 / L \gg 1$.\footnote{Using the language of \cite{Wong:2018svs}, in $X$-space the non-local coupling is manifest by a coupling of the form $\epsilon_{ij}A^+ \Psi_i^\dagger\Psi_j$, where $A^+$ is a gauge field. In the limit we consider, the gauge field vanishes.} Thus, the first term in \eqref{eq:X-spaceCurv} agrees with \eqref{eq:x-spaceCurv}. The other terms involving couplings between the different fermions are non-local in $x$-space, don't have a geometric interpretation, and vanish in the limit $b_2/L\to\infty$.

One last interesting question that we want to discuss is what information we obtain from the QEMC involving two exact generators, i.e. ${\cal R}_{a_2b_2}^Q=[V^Q_{a_2},V^Q_{b_2}]$. With the expressions for the generators of modular transport in the semi-classical theory, this is easily obtained. First, let us consider the situation without an island. In this case, there is just a single interval in the bath and modular flow is completely local. In this case, we obtain the single-interval result given by
\beq
\text{No Island:} \qquad {\cal R}_{a_2b_2}^Q = {\cal R}_{a_2b_2} = \frac{2}{(a_2-b_2)^2}Q_{11}^{(1)} ~.
\eeq
This expression is local, since it only involves the fermion $\Psi_1$. We recognize the prefactor as the coefficient of the modular curvature of a single interval in the bulk.

Now consider the situation with an island present. In that case, we can make use of the expression of the generators in $X$-space to find\footnote{The commutators $[V_{a_i},V_{a_j}] = [V_{b_i},V_{b_j}]=0$.}
\beq
{\cal R}_{a_2b_2}^Q = [V_{a_2},V_{b_2}] + \left(\frac{\partial b_1}{\partial a_2}\right)\left(\frac{\partial a_1}{\partial b_2}\right)[V_{b_1},V_{a_1}] ~.
\eeq
The general result looks rather complicated. Expanding for $a_2 / L \ll 1$ and $b_2 / L \gg 1$ we find that the leading order term in the expansion vanishes. Because we saw that in this limit modular flow becomes local in the diamond $x=[-L,0]$, the vanishing of the leading order term implies that this component of the QEMC contains purely non-local information. The first non-zero order in the expansion is ${\cal O}(L/b_2)^{3/2}$. We find
\beq \label{eq:X-spaceCurv2}
\text{With Island:} \qquad {\cal R}_{a_2b_2}^Q = -\frac1{\sqrt{8b_2^3L}}\left(Q_{12}^{(0)}-Q_{21}^{(0)} - 6\left(Q_{12}^{(1)}+Q_{21}^{(1)}\right) + 2Q_{21}^{(2)}\right) ~.
\eeq
Indeed, this term is purely non-local and represents non-geometric information. We therefore see that ${\cal R}_{a_2b_2}^Q$ only has a local interpretation in the absence of an island and becomes non-local whenever a non-trivial QES contributes.

\section{Discussion} \label{sec:Discussion}
In this paper, we studied modular transport in JT gravity coupled to a non-gravitational bath in the presence of an island. Due to the non-trivial island, performing modular flow on this system involves two disjoint intervals: one in the gravitating bulk and one in the non-gravitating bath. In order to use analytical expressions, we coupled the JT theory to a two-dimensional free fermion theory for which the exact modular Hamiltonian was derived in \cite{Casini:2009vk}. Using a two-dimensional version of the JLMS formula---including the possibility of a non-trivial quantum extremal surface---we were able to define the `fine-grained' generators of modular transport associated to an interval in the non-gravitional bath region. In the semi-classical theory these generators act on both the bulk and bath interval. We computed the modular curvature from these generators, which we refer to as the Quantum Extremal Modular Curvature, or QEMC.

The QEMC is non-local in general, but in a certain simplifying limit reduces to the known (local) result for a single interval in the bulk. In this limit, the QEMC probes the bulk Riemann curvature of this interval in a similar fashion as previous studies \cite{Czech:2019vih,DeBoer:2019kdj} of the modular curvature in holography without islands. The upshot of our approach, however, is that it probes the Riemann curvature in a region that can be far away from the boundary. Thus, an observer in the bath can use this approach to obtain geometrical information about far-separated regions, possibly behind horizons. 

While we performed our analysis in the context of a low-dimensional model with many simplifying features in order to have analytical control, we expect this upshot to remain true more generally: it simply relies on the fact that the location of the island is tied to the location of the interval in the bath region, by virtue of the QES condition. Thus, performing transport with respect to the bath interval necessarily results in transport of the island, which can be used to probe the local curvature in that region. In particular, there is no obvious obstruction to applying our method to a more complicated model of an evaporating black holes, such as the one studied in \cite{Penington:2019npb,Almheiri:2019psf}, for which the island lies behind a horizon. Coupling the JT solution to the free fermion model, the modular Hamiltonian is given by the same expression we studied in this paper.

Although possible in principle, the modular transport protocol we described might be difficult to perform in practice. An observer would need to be able to compute the fine-grained generator of modular transport: a complicated and non-perturbative operator. It has a simple semi-classical bulk interpretation, however, in terms of a non-local generator that acts on both intervals. The challenge of constructing the fine-grained generator is the same challenge hindering an observer who tries to verify that the Hawking radiation of an evaporating black hole follows a Page curve. Similarly, the fine-grained entropy is a difficult observable to measure, but has a straightforward interpretation in the semi-classical theory. 

Away from a special OPE limit, the QEMC is non-local in the presence of an island and this reflects the non-local nature of modular flow of disjoint intervals. This is most easily seen in $X$-space where the non-local contributions in \eqref{eq:X-spaceCurv} and \eqref{eq:X-spaceCurv2} show up as a coupling between two species of fermions, $\Psi_1$ and $\Psi_2$.

The QEMC therefore has radically different behavior with or without the contribution of an island. An observer in the bath region could start with a (small) interval whose entanglement wedge does not  include an island and obtain a local curvature. Increasing the size of the interval, at some point an island will contribute and the QEMC suddenly becomes non-local. We note that this situation bears resemblance to the suggested breakdown of semi-classical physics in the presence of islands \cite{Bousso:2023kdj,Banks:2024imv} and it would be interesting to understand if islands are an avatar for a breakdown of EFT more generally.

One possible way to explore the non-local aspects of the QEMC is by considering the role of the additional contributing zero modes within the framework of von Neumann algebras. Ref. \cite{Banerjee:2023eew} considered the implications of a non-modular geometric phases for the algebra type (especially regards to the existence of a trace functional), and the reformulation of modular Berry transport in the language of von Neumann algebras will be explored in \cite{vNBerry}. Perhaps this language can suggest a semiclassical bulk interpretation for the non-local contributions to the QEMC, along the lines of \cite{Engelhardt:2023xer}. Another potential approach is to make use of the relation between modular transport and kinematic space. For single intervals in a holographic CFT, the modular curvature equals the Riemann curvature of kinematic space. Defining kinematic space in the presence of a QES, something considered in \cite{Gong:2023vuh}, can perhaps be instructive to better understand the QEMC.

Finally, although we only explicitly studied the QEMC for one JT gravity model on an AdS$_2$ background, our approach generalizes to non-AdS spacetimes as well. The assumption of our computation was the JLMS relation and did not necessarily require AdS holography. One particular interesting application would be to study modular transport in de Sitter space, similar to studies of the entanglement entropy in cosmological spacetimes \cite{Chen:2020tes,Hartman:2020khs,Sybesma:2020fxg,Balasubramanian:2020xqf,Geng:2021wcq,Aalsma:2021bit,Aguilar-Gutierrez:2021bns,Langhoff:2021uct,Kames-King:2021etp,Bousso:2022gth,Espindola:2022fqb,Svesko:2022txo,Levine:2022wos,Azarnia:2022kmp,Goswami:2022ylc,Aalsma:2022swk}.

\subsection*{Acknowledgements}
It is a pleasure to thank Jan de Boer, Bowen Chen, Bart\l omiej  Czech, Ricardo Esp\'{i}ndola, Jeremy van der Heijden, Bahman Najian and Ronak Soni for helpful discussions. We'd like to thank the Walter Burke Institute for Theoretical Physics at Caltech for hospitality while parts of this work were completed. CZ would also like to thank Arizona State University for hospitality during a semester leave. LA and CK are supported by the Heising-Simons Foundation under the “Observational Signatures of Quantum Gravity” collaboration grant 2021-2818, and CZ acknowledges a Heising-Simons Fellowship as part of this collaboration. CK is also supported by the U.S. Department of Energy under grant number DE-SC0019470. CZ is supported by a UMD Higholt Professorship.

\appendix

\section{Riemann Curvature from Fermi Normal Coordinates}\label{sec:FNC}
In this Appendix, we show how for modular Hamiltonians that reduce to boost Killing vectors only near the edge of the entanglement wedge we can use Fermi normal coordinates to obtain the Riemann curvature from the modular curvature. For simplicity we focus on three dimensions, the higher-dimensional case is also discussed in \cite{DeBoer:2019kdj}. We essentially follow their discussion, but specified to the three-dimensional case.

\subsection{General Procedure}
We consider a spacetime described by Fermi normal coordinates $X^\mu = (X^A,Y)$, where $Y$ is the direction along the geodesic. The metric is given by
\beq
\bal
g_{AB} &= \eta_{AB} - \frac13R_{ACBD}X^CX^D + {\cal O}(X^3) ~. \\
g_{AY} &= \frac23R_{YBCA}X^BX^C + {\cal O}(X^3) ~, \\
g_{YY} &= 1 - R_{YAYB}X^AX^B + {\cal O}(X^3) ~.
\eal
\eeq
The modular Hamiltonian is an approximate boost Killing vector near the edge of the entanglement wedge. Using lightcone coordinates $X^\pm = X^0 \pm X^1$ we can write this vector as
\beq
\zeta = X^+\partial_+ - X^-\partial_- + {\cal O}(X^2) ~. 
\eeq
Now consider a generator of modular transport $V_\lambda$. In the main body, we saw that it can be expressed as
\beq
V_\lambda = [K'_\lambda,K] ~,
\eeq
where $K = 2\pi\zeta$ and $K'_\lambda$ is $1/(2\pi)^2$ times the change in the modular Hamiltonian, defined analogously to \eqref{eq:Kprime}. Since $K$ and $K'_\lambda$ are both (approximate) Killing vectors, so is $V_\lambda$. This implies that it should approximately solve Killing's equation. Dropping the $\lambda$ subscript to ease notation:
\beq
\bal
g_{\nu\rho}\nabla_\mu V^\rho + g_{\mu\rho}\nabla_\nu V^\rho = {\cal O}(X) ~. \\
\eal
\eeq
Denoting the location of the new extremal surface by $X'{}^A(Y) = 0$, in Fermi normal coordinates this is solved by
\beq
\bal
V^A &=  X'{}^A(Y) + {\cal O}(X^2) ~, \\
V^Y &=  -\eta_{AB}(\partial_YX'{}^A(Y))X^B + {\cal O}(X^2) ~.
\eal
\eeq
We now split the generator as $V=V_{(+)}+V_{(-)}$:
\beq
\bal
V_{(+)} &= X'{}^+(Y)\partial_+ +\frac12X^-(\partial_Y X'{}^+(Y))\partial_Y + {\cal O}(X^2) ~, \\
V_{(-)} &= X'{}^-(Y)\partial_- +\frac12X^+(\partial_Y X'{}^-(Y))\partial_Y + {\cal O}(X^2) ~.
\eal
\eeq
With the expansion of the generators of modular transport the modular curvature is given by
\beq
{\cal R}_{+-} = [V_{(+)},V_{(-)}]~.
\eeq
However, acting with either $V_{(+)}$ or $V_{(-)}$ changes the location of the extremal surface and therefore changes the Fermi normal coordinate gauge in which the derived expressions for $V_{(+)}$ and $V_{(-)}$ are valid. As explained in \cite{DeBoer:2019kdj}, we can correct for this by adjusting the ${\cal O}(X^2)$ terms to preserve the gauge after acting with one of the generators. This results in
\beq
\bal
V^A &=  X'{}^A(Y) + \frac13X'{}^C(Y)\eta^{AD}R_{B(CD)E}X^BX^E ~, \\
V^Y &=  -\eta_{AB}(\partial_YX'{}^A(Y))X^B - \frac13X'{}^B(Y)R^Y_{\,\,\,ABC}X^AX^C  ~.
\eal
\eeq
We then find\footnote{This corrects a factor of $1/3$ absent in (3.20) of \cite{DeBoer:2019kdj}.}
\beq \label{eq:GenModCurv}
{\cal R}_{+-} = [V_{(+)},V_{(-)}] = \frac13X'{}^+X'{}^-R_{+-}^{~~~~\mu}{}_{A}X^A\partial_\mu - \frac12(\partial_YX'{}^-)(\partial_YX'{}^+)(X^+\partial_+ - X^-\partial_-) ~.
\eeq
Thus, we explicitly see how the modular curvature is related to the Riemann curvature.

\subsection{Example: Diagonal Geodesic in AdS$_3$}
We start from the AdS$_3$ metric in global coordinates:
\beq
\rmd s^2 = \ell^2(-\cosh^2\rho \,\rmd \tau^2 + \rmd\rho^2 + \sinh^2\rho\,\rmd\theta^2) ~.
\eeq
We want to construct Fermi normal coordinates along the diagonal geodesic with endpoints $\tau=0$ and $\theta=[0,\pi]$. For this purpose, it is useful to first consider Poincaré coordinates which are naturally adapted to this geodesic:
\beq
\rmd s^2 = \frac{\ell^2}{y^2}(-\rmd t^2 + \rmd x^2 + \rmd y^2) ~,
\eeq
where we define the coordinate system in such a way that the diagonal geodesic is parametrized by $(t,x,y) = (0,0,y(q))$, where $q$ parametrizes the geodesic. The explicit relation between these coordinates is 
\beq
\bal
t &= \frac{\ell \cosh\rho\sin\tau}{\cosh\rho\cos\tau-\sinh\rho\sin\theta} ~, \\
x &= \frac{\ell \sinh\rho\cos\theta}{\cosh\rho\cos\tau-\sinh\rho\sin\theta} ~, \\
y &= \frac{\ell}{\cosh\rho\cos\tau-\sinh\rho\sin\theta} ~.
\eal
\eeq
The explicit coordinate transformation to Fermi normal coordinates $(T,X,Y)$ is now given by:
\beq
\bal
t &= e^{-Y/\ell}T\left(1+\frac{T^2-X^2}{3\ell^2}\right)~,\\
x &= e^{-Y/\ell}X\left(1+\frac{T^2-X^2}{3\ell^2}\right)~,\\
y &= e^{-Y/\ell}\ell\left(1+\frac{T^2-X^2}{2\ell^2}\right)~.
\eal
\eeq
The geodesic is described by $(T,X,Y)=(0,0,Y(q))$. Expanding the metric up to second order in $(T,X)$ away from the geodesic we find
\beq \label{eq:FNCmetric}
\rmd s^2 = -\left(1+\frac{X^2}{3\ell^2}\right)\rmd T^2 + \left(1-\frac{T^2}{3\ell^2}\right)\rmd X^2 + \left(1-\frac{T^2-X^2}{\ell^2}\right)\rmd Y^2 + \frac{2TX}{3\ell^2}\rmd T\rmd X ~.
\eeq
This indeed takes the form of Fermi normal coordinates. Denoting $X^A = (T,X)$ we have
\beq
\bal
g_{YY} &= 1 - R_{YAYB}X^AX^B + {\cal O}(X^3) ~,\\
g_{YA} &= -\frac23 R_{YBAC}X^BX^C + {\cal O}(X^3) ~, \\
g_{AB} &= \eta_{AB} - \frac13R_{ACBD}X^CX^D + {\cal O}(X^3) ~.
\eal
\eeq
We now have to consider a particular deformation of the diagonal geodesic. We choose to consider a rotation in global coordinates, given by $V=\partial_\theta$. It is easy to compute that in Fermi normal coordinates, such a rotation corresponds to the following location of the new extremal surface in terms of the $X^\pm = T \pm X$ coordinates:
\beq
X'{}^\pm(Y) = \pm \ell \sinh(Y/\ell) ~.
\eeq
Evaluating the Riemann components in Fermi normal coordinates we find that up to ${\cal O}(X^2)$ the generators are given by
\beq
\bal
V_{(+)} &= +\ell\sinh(Y/\ell)\left(1-\frac{X^+(X^++X^-)}{6\ell^2}\right)\partial_+ + \frac{X^-}{2}\cosh(Y/\ell)\partial_Y ~,\\
V_{(-)} &= -\ell\sinh(Y/\ell)\left(1-\frac{X^-(X^++X^-)}{6\ell^2}\right)\partial_- - \frac{X^+}{2}\cosh(Y/\ell)\partial_Y ~.\\
\eal
\eeq
Computing the commutator we find
\beq
{\cal R}_{+-} = \frac{X^+}{6}\left(2+\cosh\left(2Y/\ell\right)\right)\partial_+ - \frac{X^-}{6}\left(2+\cosh\left(2Y/\ell\right)\right)\partial_- ~.
\eeq
This indeed matches the general expression \eqref{eq:GenModCurv} when evaluated on the metric \eqref{eq:FNCmetric}.

\section{Generators of Modular Transport}\label{sec:generators}
Here we write down the generators of modular transport both in $x$-space as well as in $X$-space as derived in \cite{Chen:2022nwf}.

\subsection{Generators in $x$-space}
For a set of two disjoint intervals described by $[a_1,b_1]\cup[a_2,b_2]$ the generators of modular transport are given by\footnote{As in the main body, we focus on the $+$ component and drop the explicit $+$ superscripts. We note that our definition of the generators differs from \cite{Chen:2022nwf} by a minus sign, i.e. $V_{\lambda}^{\rm us} = -V_{\lambda}^{\rm them}$. }
\beq \label{eq:ExplicitGenerators}
\bal
&V_{a_i} = \\
&\int\rmd x\frac{\Pi_b^2(x)}{\Delta(x)^2}\left(-\frac{\Pi_a^2(x)}{(a_i-x)^2}T(x) -\frac{\xi\Pi_a(x)}{\Pi_b(a_i)\gamma(x)}\tilde T(x) + \frac{\xi(\Delta(x)-\frac{2\Pi_b(a_i)\Pi_a(x)}{(a_i-x)})}{2\Pi_b(a_i)\gamma(x)^2}\psi^\dagger(x)\psi(\tilde x)\right) ~, \\
&V_{b_i} = \\
&\int\rmd x\frac{\Pi_a^2(x)}{\Delta(x)^2}\left(-\frac{\Pi_b^2(x)}{(b_i-x)^2}T(x) +\frac{\xi\Pi_b(x)}{\Pi_a(b_i)\gamma(x)}\tilde T(x) -\frac{\xi(\Delta(x)-\frac{2\Pi_a(b_i)\Pi_b(x)}{(b_i-x)})}{2\Pi_a(b_i)\gamma(x)^2}\psi^\dagger(x)\psi(\tilde x)\right) ~.
\eal
\eeq
The operator $\tilde T$ and coefficients are defined as
\beq
\tilde T(x) = \frac12\left(\partial_x\psi^\dagger(x)\psi(\tilde x)-(\partial_x\tilde x)\psi^\dagger(x)\tilde \partial_x\psi(\tilde x)\right) ~,
\eeq
and
\beq
\bal
\Delta(x) &= (b_1+b_2-a_1-a_2)x^2 + 2(a_1a_2-b_1b_2)x + (a_1+a_2)b_1b_2-(b_1+b_2)a_1a_2 ~, \\
\gamma(x) &= (b_1+b_2-a_1-a_2)x + a_1a_2 - b_1b_2 ~, \\
\Pi_a(x) &= (a_1-x)(a_2-x) ~, \\
\Pi_b(x) &= (b_1-x)(b_2-x) ~, \\
\xi &=-(b_1-a_1)(b_1-a_2)(b_2-a_1)(b_2-a_2) ~. 
\eal
\eeq
The generators obey
\beq
[V_{a_i},H] = \partial_{a_i}H ~, \qquad [V_{b_i},H] = \partial_{b_i}H ~.
\eeq

\subsection{Generators in $X$-space}
In $X$-space the four generators $V_{\lambda=(a_1,a_2,b_1,b_2)}$ can be compactly written as
\beq
V_\lambda = \int \rmd X \sum_{ij}A_{ij}^{(\lambda)}(X)T_{ij}(X) ~.
\eeq
Defining
\beq
\tau_{a_1} = \frac{(b_1-a_2)(b_2-a_2)}{(b_1-a_1)(b_2-a_1)} ~, \qquad \tau_{b_1} = \frac{(b_2-a_1)(b_2-a_2)}{(b_1-a_1)(b_1-a_2)}~,
\eeq
the coefficients are given by the four matrices
\beq
A_{ij}^{(a_1)} = 
\begin{pmatrix}
\frac{b_2-a_2}{(b_2-b_1)(b_2-a_1)} & -\frac{\sqrt{-\tau_{a_1}}}{b_2-b_1} \\
-\frac{\sqrt{-\tau_{a_1}}}{b_2-b_1} & \frac{b_1-a_2}{(b_1-b_2)(b_1-a_1)}
\end{pmatrix}
~, \qquad
A_{ij}^{(a_2)} = 
\begin{pmatrix}
\frac{b_2-a_1}{(b_2-a_2)(b_2-b_1)} & -\frac{\sqrt{-\tau_{a_2}}}{b_1-b_2} \\
-\frac{\sqrt{-\tau_{a_2}}}{b_1-b_2} & \frac{b_1-a_1}{(b_1-b_2)(b_1-a_2)}
\end{pmatrix} ~,
\eeq
and
\beq
A_{ij}^{(b_1)} =
\begin{pmatrix}
0 & 0 \\
0 & -\frac{(b_1-b_2)}{(b_1-a_1)(b_1-a_2)}X^2
\end{pmatrix}
~, \qquad
A_{ij}^{(b_2)} =
\begin{pmatrix}
-\frac{(b_2-b_1)}{(b_2-a_1)(b_2-a_2)}X^2 & 0 \\
0 & 0
\end{pmatrix} ~.
\eeq

\bibliographystyle{utphys}
\bibliography{refs}

\providecommand{\href}[2]{#2}\begingroup\raggedright\begin{thebibliography}{10}

\bibitem{Ryu:2006bv}
S.~Ryu and T.~Takayanagi, ``{Holographic derivation of entanglement entropy
  from AdS/CFT},'' \href{http://dx.doi.org/10.1103/PhysRevLett.96.181602}{{\em
  Phys. Rev. Lett.} {\bfseries 96} (2006) 181602},
  \href{http://arxiv.org/abs/hep-th/0603001}{{\ttfamily arXiv:hep-th/0603001}}.

\bibitem{Hubeny:2007xt}
V.~E. Hubeny, M.~Rangamani, and T.~Takayanagi, ``{A Covariant holographic
  entanglement entropy proposal},''
  \href{http://dx.doi.org/10.1088/1126-6708/2007/07/062}{{\em JHEP} {\bfseries
  07} (2007) 062}, \href{http://arxiv.org/abs/0705.0016}{{\ttfamily
  arXiv:0705.0016 [hep-th]}}.

\bibitem{Czech:2017zfq}
B.~Czech, L.~Lamprou, S.~Mccandlish, and J.~Sully, ``{Modular Berry Connection
  for Entangled Subregions in AdS/CFT},''
  \href{http://dx.doi.org/10.1103/PhysRevLett.120.091601}{{\em Phys. Rev.
  Lett.} {\bfseries 120} no.~9, (2018) 091601},
  \href{http://arxiv.org/abs/1712.07123}{{\ttfamily arXiv:1712.07123
  [hep-th]}}.

\bibitem{Czech:2019vih}
B.~Czech, J.~De~Boer, D.~Ge, and L.~Lamprou, ``{A modular sewing kit for
  entanglement wedges},'' \href{http://dx.doi.org/10.1007/JHEP11(2019)094}{{\em
  JHEP} {\bfseries 11} (2019) 094},
  \href{http://arxiv.org/abs/1903.04493}{{\ttfamily arXiv:1903.04493
  [hep-th]}}.

\bibitem{Czech:2015qta}
B.~Czech, L.~Lamprou, S.~McCandlish, and J.~Sully, ``{Integral Geometry and
  Holography},'' \href{http://dx.doi.org/10.1007/JHEP10(2015)175}{{\em JHEP}
  {\bfseries 10} (2015) 175}, \href{http://arxiv.org/abs/1505.05515}{{\ttfamily
  arXiv:1505.05515 [hep-th]}}.

\bibitem{deBoer:2015kda}
J.~de~Boer, M.~P. Heller, R.~C. Myers, and Y.~Neiman, ``{Holographic de Sitter
  Geometry from Entanglement in Conformal Field Theory},''
  \href{http://dx.doi.org/10.1103/PhysRevLett.116.061602}{{\em Phys. Rev.
  Lett.} {\bfseries 116} no.~6, (2016) 061602},
  \href{http://arxiv.org/abs/1509.00113}{{\ttfamily arXiv:1509.00113
  [hep-th]}}.

\bibitem{Czech:2016xec}
B.~Czech, L.~Lamprou, S.~McCandlish, B.~Mosk, and J.~Sully, ``{A Stereoscopic
  Look into the Bulk},'' \href{http://dx.doi.org/10.1007/JHEP07(2016)129}{{\em
  JHEP} {\bfseries 07} (2016) 129},
  \href{http://arxiv.org/abs/1604.03110}{{\ttfamily arXiv:1604.03110
  [hep-th]}}.

\bibitem{deBoer:2016pqk}
J.~de~Boer, F.~M. Haehl, M.~P. Heller, and R.~C. Myers, ``{Entanglement,
  holography and causal diamonds},''
  \href{http://dx.doi.org/10.1007/JHEP08(2016)162}{{\em JHEP} {\bfseries 08}
  (2016) 162}, \href{http://arxiv.org/abs/1606.03307}{{\ttfamily
  arXiv:1606.03307 [hep-th]}}.

\bibitem{Nogueira:2021ngh}
F.~S. Nogueira, S.~Banerjee, M.~Dorband, R.~Meyer, J.~v.~d. Brink, and
  J.~Erdmenger, ``{Geometric phases distinguish entangled states in wormhole
  quantum mechanics},''
  \href{http://dx.doi.org/10.1103/PhysRevD.105.L081903}{{\em Phys. Rev. D}
  {\bfseries 105} no.~8, (2022) L081903},
  \href{http://arxiv.org/abs/2109.06190}{{\ttfamily arXiv:2109.06190
  [hep-th]}}.

\bibitem{Banerjee:2022jnv}
S.~Banerjee, M.~Dorband, J.~Erdmenger, R.~Meyer, and A.-L. Weigel, ``{Berry
  phases, wormholes and factorization in AdS/CFT},''
  \href{http://dx.doi.org/10.1007/JHEP08(2022)162}{{\em JHEP} {\bfseries 08}
  (2022) 162}, \href{http://arxiv.org/abs/2202.11717}{{\ttfamily
  arXiv:2202.11717 [hep-th]}}.

\bibitem{Bao:2019bib}
N.~Bao, C.~Cao, S.~Fischetti, and C.~Keeler, ``{Towards Bulk Metric
  Reconstruction from Extremal Area Variations},''
  \href{http://dx.doi.org/10.1088/1361-6382/ab377f}{{\em Class. Quant. Grav.}
  {\bfseries 36} no.~18, (2019) 185002},
  \href{http://arxiv.org/abs/1904.04834}{{\ttfamily arXiv:1904.04834
  [hep-th]}}.

\bibitem{Engelhardt:2016wgb}
N.~Engelhardt and G.~T. Horowitz, ``{Towards a Reconstruction of General Bulk
  Metrics},'' \href{http://dx.doi.org/10.1088/1361-6382/34/1/015004}{{\em
  Class. Quant. Grav.} {\bfseries 34} no.~1, (2017) 015004},
  \href{http://arxiv.org/abs/1605.01070}{{\ttfamily arXiv:1605.01070
  [hep-th]}}.

\bibitem{Engelhardt:2016crc}
N.~Engelhardt and G.~T. Horowitz, ``{Recovering the spacetime metric from a
  holographic dual},''
  \href{http://dx.doi.org/10.4310/ATMP.2017.v21.n7.a2}{{\em Adv. Theor. Math.
  Phys.} {\bfseries 21} (2017) 1635--1653},
  \href{http://arxiv.org/abs/1612.00391}{{\ttfamily arXiv:1612.00391
  [hep-th]}}.

\bibitem{DeBoer:2019kdj}
J.~De~Boer and L.~Lamprou, ``{Holographic Order from Modular Chaos},''
  \href{http://dx.doi.org/10.1007/JHEP06(2020)024}{{\em JHEP} {\bfseries 06}
  (2020) 024}, \href{http://arxiv.org/abs/1912.02810}{{\ttfamily
  arXiv:1912.02810 [hep-th]}}.

\bibitem{deBoer:2021zlm}
J.~de~Boer, R.~Esp\'\i{}ndola, B.~Najian, D.~Patramanis, J.~van~der Heijden,
  and C.~Zukowski, ``{Virasoro entanglement Berry phases},''
  \href{http://dx.doi.org/10.1007/JHEP03(2022)179}{{\em JHEP} {\bfseries 03}
  (2022) 179}, \href{http://arxiv.org/abs/2111.05345}{{\ttfamily
  arXiv:2111.05345 [hep-th]}}.

\bibitem{Czech:2023zmq}
B.~Czech, J.~de~Boer, R.~Esp\'\i{}ndola, B.~Najian, J.~van~der Heijden, and
  C.~Zukowski, ``{Changing states in holography: From modular Berry curvature
  to the bulk symplectic form},''
  \href{http://dx.doi.org/10.1103/PhysRevD.108.066003}{{\em Phys. Rev. D}
  {\bfseries 108} no.~6, (2023) 066003},
  \href{http://arxiv.org/abs/2305.16384}{{\ttfamily arXiv:2305.16384
  [hep-th]}}.

\bibitem{Dong:2016eik}
X.~Dong, D.~Harlow, and A.~C. Wall, ``{Reconstruction of Bulk Operators within
  the Entanglement Wedge in Gauge-Gravity Duality},''
  \href{http://dx.doi.org/10.1103/PhysRevLett.117.021601}{{\em Phys. Rev.
  Lett.} {\bfseries 117} no.~2, (2016) 021601},
  \href{http://arxiv.org/abs/1601.05416}{{\ttfamily arXiv:1601.05416
  [hep-th]}}.

\bibitem{Balasubramanian:2014sra}
V.~Balasubramanian, B.~D. Chowdhury, B.~Czech, and J.~de~Boer, ``{Entwinement
  and the emergence of spacetime},''
  \href{http://dx.doi.org/10.1007/JHEP01(2015)048}{{\em JHEP} {\bfseries 01}
  (2015) 048}, \href{http://arxiv.org/abs/1406.5859}{{\ttfamily arXiv:1406.5859
  [hep-th]}}.

\bibitem{Freivogel:2014lja}
B.~Freivogel, R.~Jefferson, L.~Kabir, B.~Mosk, and I.-S. Yang, ``{Casting
  Shadows on Holographic Reconstruction},''
  \href{http://dx.doi.org/10.1103/PhysRevD.91.086013}{{\em Phys. Rev. D}
  {\bfseries 91} no.~8, (2015) 086013},
  \href{http://arxiv.org/abs/1412.5175}{{\ttfamily arXiv:1412.5175 [hep-th]}}.

\bibitem{Engelhardt:2015dta}
N.~Engelhardt and S.~Fischetti, ``{Covariant Constraints on Hole-ography},''
  \href{http://dx.doi.org/10.1088/0264-9381/32/19/195021}{{\em Class. Quant.
  Grav.} {\bfseries 32} no.~19, (2015) 195021},
  \href{http://arxiv.org/abs/1507.00354}{{\ttfamily arXiv:1507.00354
  [hep-th]}}.

\bibitem{Engelhardt:2014gca}
N.~Engelhardt and A.~C. Wall, ``{Quantum Extremal Surfaces: Holographic
  Entanglement Entropy beyond the Classical Regime},''
  \href{http://dx.doi.org/10.1007/JHEP01(2015)073}{{\em JHEP} {\bfseries 01}
  (2015) 073}, \href{http://arxiv.org/abs/1408.3203}{{\ttfamily arXiv:1408.3203
  [hep-th]}}.

\bibitem{Penington:2019npb}
G.~Penington, ``{Entanglement Wedge Reconstruction and the Information
  Paradox},'' \href{http://dx.doi.org/10.1007/JHEP09(2020)002}{{\em JHEP}
  {\bfseries 09} (2020) 002}, \href{http://arxiv.org/abs/1905.08255}{{\ttfamily
  arXiv:1905.08255 [hep-th]}}.

\bibitem{Almheiri:2019psf}
A.~Almheiri, N.~Engelhardt, D.~Marolf, and H.~Maxfield, ``{The entropy of bulk
  quantum fields and the entanglement wedge of an evaporating black hole},''
  \href{http://dx.doi.org/10.1007/JHEP12(2019)063}{{\em JHEP} {\bfseries 12}
  (2019) 063}, \href{http://arxiv.org/abs/1905.08762}{{\ttfamily
  arXiv:1905.08762 [hep-th]}}.

\bibitem{Casini:2009vk}
H.~Casini and M.~Huerta, ``{Reduced density matrix and internal dynamics for
  multicomponent regions},''
  \href{http://dx.doi.org/10.1088/0264-9381/26/18/185005}{{\em Class. Quant.
  Grav.} {\bfseries 26} (2009) 185005},
  \href{http://arxiv.org/abs/0903.5284}{{\ttfamily arXiv:0903.5284 [hep-th]}}.

\bibitem{Chen:2019iro}
Y.~Chen, ``{Pulling Out the Island with Modular Flow},''
  \href{http://dx.doi.org/10.1007/JHEP03(2020)033}{{\em JHEP} {\bfseries 03}
  (2020) 033}, \href{http://arxiv.org/abs/1912.02210}{{\ttfamily
  arXiv:1912.02210 [hep-th]}}.

\bibitem{Chen:2022nwf}
B.~Chen, B.~Czech, L.-Y. Hung, and G.~Wong, ``{Modular parallel transport of
  multiple intervals in 1+1-dimensional free fermion theory},''
  \href{http://dx.doi.org/10.1007/JHEP03(2023)147}{{\em JHEP} {\bfseries 03}
  (2023) 147}, \href{http://arxiv.org/abs/2211.12545}{{\ttfamily
  arXiv:2211.12545 [hep-th]}}.

\bibitem{Bousso:2023kdj}
R.~Bousso and G.~Penington, ``{Islands Far Outside the Horizon},''
  \href{http://arxiv.org/abs/2312.03078}{{\ttfamily arXiv:2312.03078
  [hep-th]}}.

\bibitem{Banks:2024imv}
T.~Banks, P.~Draper, and M.~Karydas, ``{Breakdown of Field Theory in
  Near-Horizon Regions},'' \href{http://arxiv.org/abs/2401.03572}{{\ttfamily
  arXiv:2401.03572 [hep-th]}}.

\bibitem{Bisognano:1976za}
J.~J. Bisognano and E.~H. Wichmann, ``{On the Duality Condition for Quantum
  Fields},'' \href{http://dx.doi.org/10.1063/1.522898}{{\em J. Math. Phys.}
  {\bfseries 17} (1976) 303--321}.

\bibitem{Faulkner:2017vdd}
T.~Faulkner and A.~Lewkowycz, ``{Bulk locality from modular flow},''
  \href{http://dx.doi.org/10.1007/JHEP07(2017)151}{{\em JHEP} {\bfseries 07}
  (2017) 151}, \href{http://arxiv.org/abs/1704.05464}{{\ttfamily
  arXiv:1704.05464 [hep-th]}}.

\bibitem{Balakrishnan:2020lbp}
S.~Balakrishnan and O.~Parrikar, ``{Modular Hamiltonians for Euclidean Path
  Integral States},'' \href{http://arxiv.org/abs/2002.00018}{{\ttfamily
  arXiv:2002.00018 [hep-th]}}.

\bibitem{Jafferis:2015del}
D.~L. Jafferis, A.~Lewkowycz, J.~Maldacena, and S.~J. Suh, ``{Relative entropy
  equals bulk relative entropy},''
  \href{http://dx.doi.org/10.1007/JHEP06(2016)004}{{\em JHEP} {\bfseries 06}
  (2016) 004}, \href{http://arxiv.org/abs/1512.06431}{{\ttfamily
  arXiv:1512.06431 [hep-th]}}.

\bibitem{Almheiri:2014cka}
A.~Almheiri and J.~Polchinski, ``{Models of AdS$_{2}$ backreaction and
  holography},'' \href{http://dx.doi.org/10.1007/JHEP11(2015)014}{{\em JHEP}
  {\bfseries 11} (2015) 014}, \href{http://arxiv.org/abs/1402.6334}{{\ttfamily
  arXiv:1402.6334 [hep-th]}}.

\bibitem{Engelsoy:2016xyb}
J.~Engels\"oy, T.~G. Mertens, and H.~Verlinde, ``{An investigation of AdS$_{2}$
  backreaction and holography},''
  \href{http://dx.doi.org/10.1007/JHEP07(2016)139}{{\em JHEP} {\bfseries 07}
  (2016) 139}, \href{http://arxiv.org/abs/1606.03438}{{\ttfamily
  arXiv:1606.03438 [hep-th]}}.

\bibitem{Christensen:1977jc}
S.~M. Christensen and S.~A. Fulling, ``{Trace Anomalies and the Hawking
  Effect},'' \href{http://dx.doi.org/10.1103/PhysRevD.15.2088}{{\em Phys. Rev.
  D} {\bfseries 15} (1977) 2088--2104}.

\bibitem{Fabbri:2005mw}
A.~Fabbri and J.~Navarro-Salas, {\em {Modeling black hole evaporation}}.
\newblock 2005.

\bibitem{Almheiri:2019yqk}
A.~Almheiri, R.~Mahajan, and J.~Maldacena, ``{Islands outside the horizon},''
  \href{http://arxiv.org/abs/1910.11077}{{\ttfamily arXiv:1910.11077
  [hep-th]}}.

\bibitem{Hartman:2013mia}
T.~Hartman, ``{Entanglement Entropy at Large Central Charge},''
  \href{http://arxiv.org/abs/1303.6955}{{\ttfamily arXiv:1303.6955 [hep-th]}}.

\bibitem{Barrella:2013wja}
T.~Barrella, X.~Dong, S.~A. Hartnoll, and V.~L. Martin, ``{Holographic
  entanglement beyond classical gravity},''
  \href{http://dx.doi.org/10.1007/JHEP09(2013)109}{{\em JHEP} {\bfseries 09}
  (2013) 109}, \href{http://arxiv.org/abs/1306.4682}{{\ttfamily arXiv:1306.4682
  [hep-th]}}.

\bibitem{Rolph:2021nan}
A.~Rolph, ``{Local measures of entanglement in black holes and CFTs},''
  \href{http://dx.doi.org/10.21468/SciPostPhys.12.3.079}{{\em SciPost Phys.}
  {\bfseries 12} no.~3, (2022) 079},
  \href{http://arxiv.org/abs/2107.11385}{{\ttfamily arXiv:2107.11385
  [hep-th]}}.

\bibitem{Callebaut:2018nlq}
N.~Callebaut and H.~Verlinde, ``{Entanglement Dynamics in 2D CFT with Boundary:
  Entropic origin of JT gravity and Schwarzian QM},''
  \href{http://dx.doi.org/10.1007/JHEP05(2019)045}{{\em JHEP} {\bfseries 05}
  (2019) 045}, \href{http://arxiv.org/abs/1808.05583}{{\ttfamily
  arXiv:1808.05583 [hep-th]}}.

\bibitem{Callebaut:2018xfu}
N.~Callebaut, ``{The gravitational dynamics of kinematic space},''
  \href{http://dx.doi.org/10.1007/JHEP02(2019)153}{{\em JHEP} {\bfseries 02}
  (2019) 153}, \href{http://arxiv.org/abs/1808.10431}{{\ttfamily
  arXiv:1808.10431 [hep-th]}}.

\bibitem{Pedraza:2021cvx}
J.~F. Pedraza, A.~Svesko, W.~Sybesma, and M.~R. Visser, ``{Semi-classical
  thermodynamics of quantum extremal surfaces in Jackiw-Teitelboim gravity},''
  \href{http://dx.doi.org/10.1007/JHEP12(2021)134}{{\em JHEP} {\bfseries 12}
  (2021) 134}, \href{http://arxiv.org/abs/2107.10358}{{\ttfamily
  arXiv:2107.10358 [hep-th]}}.

\bibitem{Wong:2018svs}
G.~Wong, ``{Gluing together Modular flows with free fermions},''
  \href{http://dx.doi.org/10.1007/JHEP04(2019)045}{{\em JHEP} {\bfseries 04}
  (2019) 045}, \href{http://arxiv.org/abs/1805.10651}{{\ttfamily
  arXiv:1805.10651 [hep-th]}}.

\bibitem{Banerjee:2023eew}
S.~Banerjee, M.~Dorband, J.~Erdmenger, and A.-L. Weigel, ``{Geometric phases
  characterise operator algebras and missing information},''
  \href{http://dx.doi.org/10.1007/JHEP10(2023)026}{{\em JHEP} {\bfseries 10}
  (2023) 026}, \href{http://arxiv.org/abs/2306.00055}{{\ttfamily
  arXiv:2306.00055 [hep-th]}}.

\bibitem{vNBerry}
J.~de~Boer, J.~van~der Heijden, B.~Najian, and C.~Zukowski. To appear.

\bibitem{Engelhardt:2023xer}
N.~Engelhardt and H.~Liu, ``{Algebraic ER=EPR and Complexity Transfer},''
  \href{http://arxiv.org/abs/2311.04281}{{\ttfamily arXiv:2311.04281
  [hep-th]}}.

\bibitem{Gong:2023vuh}
A.~Gong, C.-B. Chen, and F.-W. Shu, ``{Kinematic space for quantum extremal
  surface},'' \href{http://arxiv.org/abs/2305.15885}{{\ttfamily
  arXiv:2305.15885 [hep-th]}}.

\bibitem{Chen:2020tes}
Y.~Chen, V.~Gorbenko, and J.~Maldacena, ``{Bra-ket wormholes in gravitationally
  prepared states},'' \href{http://dx.doi.org/10.1007/JHEP02(2021)009}{{\em
  JHEP} {\bfseries 02} (2021) 009},
  \href{http://arxiv.org/abs/2007.16091}{{\ttfamily arXiv:2007.16091
  [hep-th]}}.

\bibitem{Hartman:2020khs}
T.~Hartman, Y.~Jiang, and E.~Shaghoulian, ``{Islands in cosmology},''
  \href{http://dx.doi.org/10.1007/JHEP11(2020)111}{{\em JHEP} {\bfseries 11}
  (2020) 111}, \href{http://arxiv.org/abs/2008.01022}{{\ttfamily
  arXiv:2008.01022 [hep-th]}}.

\bibitem{Sybesma:2020fxg}
W.~Sybesma, ``{Pure de Sitter space and the island moving back in time},''
  \href{http://dx.doi.org/10.1088/1361-6382/abff9a}{{\em Class. Quant. Grav.}
  {\bfseries 38} no.~14, (2021) 145012},
  \href{http://arxiv.org/abs/2008.07994}{{\ttfamily arXiv:2008.07994
  [hep-th]}}.

\bibitem{Balasubramanian:2020xqf}
V.~Balasubramanian, A.~Kar, and T.~Ugajin, ``{Islands in de Sitter space},''
  \href{http://dx.doi.org/10.1007/JHEP02(2021)072}{{\em JHEP} {\bfseries 02}
  (2021) 072}, \href{http://arxiv.org/abs/2008.05275}{{\ttfamily
  arXiv:2008.05275 [hep-th]}}.

\bibitem{Geng:2021wcq}
H.~Geng, Y.~Nomura, and H.-Y. Sun, ``{Information paradox and its resolution in
  de Sitter holography},''
  \href{http://dx.doi.org/10.1103/PhysRevD.103.126004}{{\em Phys. Rev. D}
  {\bfseries 103} no.~12, (2021) 126004},
  \href{http://arxiv.org/abs/2103.07477}{{\ttfamily arXiv:2103.07477
  [hep-th]}}.

\bibitem{Aalsma:2021bit}
L.~Aalsma and W.~Sybesma, ``{The Price of Curiosity: Information Recovery in de
  Sitter Space},'' \href{http://dx.doi.org/10.1007/JHEP05(2021)291}{{\em JHEP}
  {\bfseries 05} (2021) 291}, \href{http://arxiv.org/abs/2104.00006}{{\ttfamily
  arXiv:2104.00006 [hep-th]}}.

\bibitem{Aguilar-Gutierrez:2021bns}
S.~E. Aguilar-Gutierrez, A.~Chatwin-Davies, T.~Hertog, N.~Pinzani-Fokeeva, and
  B.~Robinson, ``{Islands in Multiverse Models},''
  \href{http://dx.doi.org/10.1007/JHEP05(2022)137}{{\em JHEP} {\bfseries 11}
  (2021) 212}, \href{http://arxiv.org/abs/2108.01278}{{\ttfamily
  arXiv:2108.01278 [hep-th]}}. [Addendum: JHEP 05, 137 (2022), Erratum: JHEP
  05, 082 (2022)].

\bibitem{Langhoff:2021uct}
K.~Langhoff, C.~Murdia, and Y.~Nomura, ``{Multiverse in an inverted island},''
  \href{http://dx.doi.org/10.1103/PhysRevD.104.086007}{{\em Phys. Rev. D}
  {\bfseries 104} no.~8, (2021) 086007},
  \href{http://arxiv.org/abs/2106.05271}{{\ttfamily arXiv:2106.05271
  [hep-th]}}.

\bibitem{Kames-King:2021etp}
J.~Kames-King, E.~M.~H. Verheijden, and E.~P. Verlinde, ``{No Page curves for
  the de Sitter horizon},''
  \href{http://dx.doi.org/10.1007/JHEP03(2022)040}{{\em JHEP} {\bfseries 03}
  (2022) 040}, \href{http://arxiv.org/abs/2108.09318}{{\ttfamily
  arXiv:2108.09318 [hep-th]}}.

\bibitem{Bousso:2022gth}
R.~Bousso and E.~Wildenhain, ``{Islands in closed and open universes},''
  \href{http://dx.doi.org/10.1103/PhysRevD.105.086012}{{\em Phys. Rev. D}
  {\bfseries 105} no.~8, (2022) 086012},
  \href{http://arxiv.org/abs/2202.05278}{{\ttfamily arXiv:2202.05278
  [hep-th]}}.

\bibitem{Espindola:2022fqb}
R.~Esp\'\i{}ndola, B.~Najian, and D.~Nikolakopoulou, ``{Islands in FRW
  Cosmologies},'' \href{http://arxiv.org/abs/2203.04433}{{\ttfamily
  arXiv:2203.04433 [hep-th]}}.

\bibitem{Svesko:2022txo}
A.~Svesko, E.~Verheijden, E.~P. Verlinde, and M.~R. Visser, ``{Quasi-local
  energy and microcanonical entropy in two-dimensional nearly de Sitter
  gravity},'' \href{http://dx.doi.org/10.1007/JHEP08(2022)075}{{\em JHEP}
  {\bfseries 08} (2022) 075}, \href{http://arxiv.org/abs/2203.00700}{{\ttfamily
  arXiv:2203.00700 [hep-th]}}.

\bibitem{Levine:2022wos}
A.~Levine and E.~Shaghoulian, ``{Encoding beyond cosmological horizons in de
  Sitter JT gravity},'' \href{http://dx.doi.org/10.1007/JHEP02(2023)179}{{\em
  JHEP} {\bfseries 02} (2023) 179},
  \href{http://arxiv.org/abs/2204.08503}{{\ttfamily arXiv:2204.08503
  [hep-th]}}.

\bibitem{Azarnia:2022kmp}
S.~Azarnia and R.~Fareghbal, ``{Islands in Kerr\textendash{}de Sitter spacetime
  and their flat limit},''
  \href{http://dx.doi.org/10.1103/PhysRevD.106.026012}{{\em Phys. Rev. D}
  {\bfseries 106} no.~2, (2022) 026012},
  \href{http://arxiv.org/abs/2204.08488}{{\ttfamily arXiv:2204.08488
  [hep-th]}}.

\bibitem{Goswami:2022ylc}
K.~Goswami and K.~Narayan, ``{Small Schwarzschild de Sitter black holes,
  quantum extremal surfaces and islands},''
  \href{http://dx.doi.org/10.1007/JHEP10(2022)031}{{\em JHEP} {\bfseries 10}
  (2022) 031}, \href{http://arxiv.org/abs/2207.10724}{{\ttfamily
  arXiv:2207.10724 [hep-th]}}.

\bibitem{Aalsma:2022swk}
L.~Aalsma, S.~E. Aguilar-Gutierrez, and W.~Sybesma, ``{An
  outsider\textquoteright{}s perspective on information recovery in de Sitter
  space},'' \href{http://dx.doi.org/10.1007/JHEP01(2023)129}{{\em JHEP}
  {\bfseries 01} (2023) 129}, \href{http://arxiv.org/abs/2210.12176}{{\ttfamily
  arXiv:2210.12176 [hep-th]}}.

\end{thebibliography}\endgroup

\end{document}